\documentclass{aa}

\usepackage{graphicx}
\usepackage{natbib}
\usepackage{amssymb}
\usepackage{txfonts}
\usepackage[figuresright]{rotating}
\bibpunct{(}{)}{;}{a}{}{,} 

\newcommand{\ud}{\mathrm{d}} 
\newcommand{\reff}{\mbox{$R_\mathrm{eff}$}}
\newcommand{\msun}{\mbox{$\mathrm{M_{\odot}}$}}
\newcommand{\bv}{\mbox{$B\!-\!V$}}
\newcommand{\vi}{\mbox{$V\!-\!I$}}
\newcommand{\hst}{\emph{HST}}
\newcommand{\acs}{\emph{ACS}}
\newcommand{\wfpc}{\emph{WFPC2}}

\newcommand{\ishape}{\texttt{Ishape}}

\newcommand{\kms}{\ensuremath{\mathrm{km}\,\mathrm{s}^{-1}}}
\newcommand{\bband}{\emph{F435W}}
\newcommand{\vband}{\emph{F555W}}
\newcommand{\iband}{\emph{F814W}}
\newcommand{\hband}{\emph{F658N}}

\begin{document}

\title{\emph{ACS} imaging of star clusters in M51\thanks{Based
on observations made with the NASA/ESA \emph{Hubble Space Telescope},
obtained from the data archive at the Space Telescope Institute. STScI
is operated by the association of Universities for Research in
Astronomy, Inc., under the NASA contract NAS 5-26555.}}
   \subtitle{I. Identification and radius distribution}

   \titlerunning{\emph{ACS} imaging of star clusters in M51}

   \author{R. A. Scheepmaker\inst{1} 
	\and 
	M. R. Haas\inst{1}
	\and
	M. Gieles\inst{1} 
	\and
	N. Bastian\inst{2}
	\and
	S. S. Larsen\inst{1}
	\and
	H. J. G. L. M. Lamers\inst{1}}

   \authorrunning{R. A. Scheepmaker \and M. R. Haas \and M. Gieles et al.}

\offprints{R. A. Scheepmaker, \email{scheepmaker@astro.uu.nl}}

\institute{$^1$Astronomical Institute, Utrecht University,
  Princetonplein 5, NL-3584 CC Utrecht, The Netherlands\\
  $^2$Department of Physics and Astronomy, University College London,
  Gower Street, London, WC1E 6BT, United Kingdom}

\date{Received 20 March 2007; accepted 23 April 2007}


\abstract
{Size measurements of young star clusters are valuable tools
to put constraints on the formation and early dynamical evolution of
star clusters.}
{We use \hst/\acs\ observations of the spiral galaxy M51 in
\bband, \vband\ and \iband\ to
select a large sample of star clusters with accurate effective radius
measurements in an area covering the complete disc of M51. We present the dataset
and study the radius distribution and relations between radius,
colour, arm/interarm region, galactocentric distance, mass and age.}
{We select a sample of 7698 (\bband), 6846 (\vband) and 5024 (\iband) slightly
resolved clusters and derive their effective radii (\reff) by fitting
the spatial profiles with analytical models convolved with the point
spread function. The radii of 1284 clusters are studied in detail.}
{We find cluster radii between 0.5 and $\sim$10~pc, and one
exceptionally large cluster candidate with $\reff=21.6$~pc. The median
\reff\ is 2.1~pc. We find 70 clusters in our sample which have colours
consistent with being old GC candidates and we find 6 new ``faint
fuzzy'' clusters in, or projected onto, the disc of M51. The radius
distribution can not be fitted with a power law similar to the one for
star-forming clouds. We find an increase in \reff\ with colour as well
as a higher fraction of clusters with $\bv\gtrsim 0.05$ in the
interarm regions. We find a correlation between \reff\ and
galactocentric distance ($R_{\mathrm{G}}$) of the form $\reff \propto
R_{\mathrm{G}}^{0.12\pm0.02}$, which is considerably weaker than the
observed correlation for old Milky Way GCs. We find weak
relations between cluster luminosity and radius: $\reff \propto
L^{0.15\pm0.02}$ for the interarm regions and $\reff \propto
L^{-0.11\pm0.01}$ for the spiral arm regions, but we do not observe a
correlation between cluster mass and radius.}
{The observed radius distribution indicates that shortly after the
formation of the clusters from a fractal gas, the radii of the
clusters have changed in a non-uniform way. We find tentative
evidence suggesting that clusters in spiral arms are more compact.}

\keywords{galaxies: individual: M51 -- galaxies: star clusters}

\maketitle

\section{Introduction}
  \label{sec:Introduction}
 
One of the most striking questions related to star cluster formation
concerns the transition from the densest parts of the star-forming
Giant Molecular Clouds (GMCs) to the clusters that emerge from them.
Observations show that the masses and radii of clouds follow a clear
relation of the form $R_{\mathrm{GMC}} \propto M_{\mathrm{GMC}}^{1/2}$
\citep[e.g.][]{larson81, solomon87} down to the scale of the star
forming clumps \citep{williams95}. Such a relation is expected for
self-gravitating clouds, which are in virial equilibrium and pressure
bounded by their surroundings \citep[e.g.][]{elmegreen89}. The same
mass-radius relation is observed for \emph{complexes} of multiple
clusters \citep{elmegreen99, elmegreen01a, bastian05b}. Elliptical
galaxies and very massive ($> 10^{6.5}~\msun$) stellar clusters follow
a similar relation ($R \propto M^{0.62}$ \citep{hasegan05}), which
advocates a possibility of forming very massive clusters by merging of
low mass clusters \citep{fellhauer05, kissler-patig06}.

However, for individual star clusters that emerge from the star
forming clouds/clumps, a relation between mass and radius is not
present \citep{vandenbergh91, bastian05a, jordan05} or at least
strongly weakened \citep{zepf99, hunter03, mackey03a, larsen04b,
lee05}.  Since this is already the case for very young clusters, it
indicates that during, or shortly after ($< 10$~Myr) the transition
from clouds to clusters the mass and/or the radius of the objects
change.

 These changes in mass and/or radius are likely to be reflected in
changes in the mass and radius \emph{distributions} \citep{ashman01}.
On the one hand, however, the mass distributions of both clouds and
clusters show great similarities. Both can be approximated by power
laws of the form $N(M)\ud M \propto M^{-\alpha}\ud M$, with the index
$\alpha$ in the range of 1.5--2.0 (see \citealt{sanders85, solomon87,
harris94, brand95, elmegreen96a, fukui01} for clouds and see e.g.\
\citealt{zhang99, ashman01, bik03, degrijs03a, hunter03} for
clusters). Recently, some studies have found evidence for an upper
mass truncation of the cluster mass distribution \citet{gieles06a,
dowell06}, which is also found for the mass distributions of GMCs
\citep{rosolowsky05}. On the other hand, the radius distributions are
less well constrained, especially for extra-galactic star clusters.
If we approximate the radius distributions by a power law of the form
$N(R)\ud R \propto R^{-\eta}\ud R$, the average value of the index for
GMCs is observed to be $\eta = 3.3\pm 0.3$ \citep{harris94,
elmegreen96a}, consistent with the gas having a fractal structure with
a fractal dimension $\eta-1 = 2.3\pm 0.3$ \citep{mandelbrot83,
elmegreen01b}. For young clusters in NGC~3256 \citet{zepf99} find
$\eta \approx 3.4$, while Bastian et al.\ (2005, from now on referred
to as ``B05'') find $\eta = 2.2\pm0.2$ for star clusters in the disc
of M51. This difference, however, seems to be caused by the erroneous
addition of $+1$ to the index in the result of
\citet{ashman01}.\footnote{Fitting a function of the form
$\mathrm{log}(N(r)) = a+b\cdot\mathrm{log}(r)$ results in
$b=-\eta$. However, using logarithmic binning, one fits
$\mathrm{log}(N(\mathrm{log}(r))) = a+b\cdot\mathrm{log}(r)$, in which
$b=1-\eta$. This extra term $+1$ can easily cause confusion when
comparing different distributions. Also see B05 and
\citet{elmegreen96a}.}
    
Our understanding of whether or not the mass and radius distribution
of clouds and clusters are similar, and to which degree, is directly
coupled to our understanding of star formation and the early evolution
of star clusters.  Besides that, possible explanations for the lack of
a mass-radius relation for clusters which change the mass and/or the
radius of the clusters in a non-uniform way, are likely to affect the
mass and radius distributions \citep{ashman01}. It is therefore
important to put better constraints on these distributions, and in
this work we will focus on the radius distribution of young star
clusters.

In the study presented here we exploit the superb resolution and large
field-of-view of the new \hst\ \emph{Advanced Camera for Surveys}
(\acs) observations of M51, taken as part of the Hubble Heritage
Project.  These observations allow us to measure the radii of a large
sample of clusters in an area covering, for the first time, the complete
disc of M51 and its companion, NGC~5195 at a $0.05\arcsec$ resolution.
In this work, which is the first in a series of papers, we
present the dataset and we study the radius distribution for the
complete cluster sample and for cluster samples with a different
background surface brightness (``background regions''). The
differences in background regions are likely to reflect differences in
environmental conditions, which could have an impact on the early
evolution of star clusters.

The radii of star clusters in M51 have already been studied by B05 and
\citet{lee05}.  However, these studies used lower resolution
\hst/\wfpc\ data and were not covering the complete disc.  Besides
this, we use different selection criteria for the clusters than B05,
based on the clusters actually being resolved and clearly separated
from nearby contaminating sources.  In combination with the larger
field-of-view and the higher resolution of the \acs\ data, this leads
to a larger sample of clusters, divided in different background
regions.

The structure of this paper is as follows: in
\S~\ref{sec:Observations, source selection and photometry} we describe
the dataset, the selection of sources and background regions and the
photometry. The radius measurements are described in
\S~\ref{sec:Radius measurements} and in \S~\ref{sec:Artificial cluster
experiments} we describe experiments with artificial clusters to
determine the accuracy and detection limits of our methods. Our
selection criteria and a description of our cluster sample are
presented in \S~\ref{sec:Selection of the sample}, followed by a
comparison between \acs\ and \wfpc\ data in \S~\ref{sec:A comparison
between ACS and WFPC2}. The radius distribution is presented in
\S~\ref{sec:Radius distribution}, and we search for correlations
between mass, radius and galactocentric distance in
\S~\ref{sec:Correlations between mass, radius and distance}. In
\S\ref{sec:Summary and conclusions} we finish with the summary and
conclusions.

\section{Observations, source selection and photometry}
  \label{sec:Observations, source selection and photometry}

\subsection{Observations}
  \label{subsec:Observations}

We make use of the \hst/\acs\ dataset of M51 (NGC~5194, a late type
Sbc galaxy), taken as part of the Hubble Heritage Project in January
2005 (proposal ID 10452, PI: S.~V.~W.~Beckwith).  The dataset consists
of 6 \acs\ pointings using the Wide Field Channel (\emph{WFC}) in
\bband\ ($\sim B$), \vband\ ($\sim V$), \iband\ ($\sim I$) and \hband\
($\sim H\alpha$), with 4 dithered exposures per filter, as is
summarized in Table~\ref{tab:Exposure}. The observations
were reduced and drizzled into one mosaic image by \citet{mutchler05}.
In summary, the standard \acs\ pipeline (CALACS) was used for bias,
dark and flat-field corrections of the individual dithered images.
The corrected images were then combined into one mosaic image using
the \texttt{MultiDrizzle} task \citep{fruchter02, koekemoer02}, which
also corrects for filter-specific geometric distortion, cosmic rays
and bad pixels.  For a complete description of the dataset we refer to
\citet{mutchler05} and the M51 mosaic website
(\texttt{http://archive.stsci.edu/prepds/m51/}). For details on the
standard pipeline processing we refer to the \emph{ACS Data Handbook}
\citep{pavlovsky05}.

The resulting mosaic image covers a region of $430\times610\arcsec$
($=17.5\times24.8$~kpc) with a resolution of 2.0~pc per pixel, where
we assumed a distance modulus of $m-M = 29.62$ from
\citet{feldmeier97}, i.e.\ a distance of $8.4\pm 0.6$~Mpc.  The
covered region is large enough to include the complete spiral disc of
M51, as well as its companion NGC~5195 (a dwarf barred spiral of early
type SB0), while at the same time the resolution is good enough to
\emph{resolve} stellar cluster candidates, i.e. to distinguish them
from stars by measuring their sizes and comparing these to the
\emph{Point Spread Function} (PSF) of the \hst/\acs\ camera.
 
\begin{table}
\caption[]{Overview of the filters and exposure times used. The
limiting magnitude is for a point source with S/N = 5.}
\label{tab:Exposure}
\begin{center}
\begin{tabular}{ccc}
\hline\hline
Filter & Exposure time & Limiting magnitude \\
\hline 
\bband & $4\times 680$~s = 2720~s & 27.3 m$_\mathrm{B}$ \\
\vband & $4\times 340$~s = 1360~s & 26.5 m$_\mathrm{V}$ \\
\iband & $4\times 340$~s = 1360~s & 25.8 m$_\mathrm{I}$ \\
\hband$+ N$[II] & $4\times 680$~s = 2720~s & -- \\
\hline\\ \\
\end{tabular}
\end{center}
\end{table}

\subsection{Source selection}
  \label{subsec:Source selection}

For source selection we used the SExtractor package \citep[][version
2.3.2]{bertin96}.  SExtractor first generates a background map by
computing the mean and standard deviation of every section of the
image with a user defined grid size for which we choose $10\times10$
pixels.  In every section the local background histogram is clipped
iteratively until every remaining pixel value is within $\pm 3\sigma$
of the median value.  The mean of the clipped histogram is then taken
as the local background value.  Every area of at least three adjacent
pixels that exceeded the background by at least $5\sigma$ was called a
source.  For details on the background estimation and the source
selection we refer to the SExtractor user manual \citep{bertin96} or
\citet{holwerda05}.  The coordinates of the sources in \bband\,
\vband\ and \iband\ were matched and only sources that were detected
in all three filters within two pixel uncertainty were kept.  This
resulted in a list of $75\,436$ sources, including cluster candidates
but also many stars and background galaxies.  We did not apply any
selection criteria based on shape, sharpness or size during the
source selection with SExtractor.  However, in \S~\ref{sec:Selection
of the sample} we use individual radii measurements to select a large
sample of cluster candidates from the source list.

\subsection{Background selection}
  \label{subsec:Background selection}

\begin{figure}
\centering
\includegraphics[width=85mm]{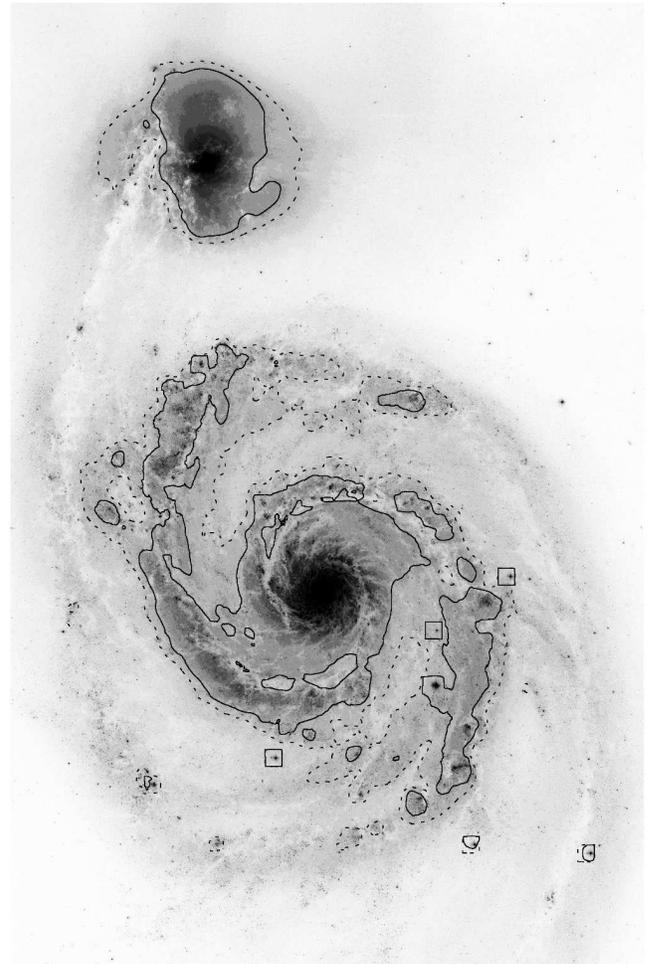}
\caption{The contours outlining the three background regions,
superimposed on the \hst/\acs\ mosaic image in the \vband\
passband. The solid lines enclose ``high background'' regions, and
everything outside the dashed lines indicates the ``low background''
regions. The regions in between solid and dashed lines are called
``intermediate background''.}
\label{fig:Backgroundregions}
\end{figure}

In order to study any possible relations between radius and
environment, and to perform completeness and accuracy tests for
different background levels, we divided the image in three regions
according to the background surface brightness. These background regions
were selected by smoothing the \vband\ image with a boxcar average of
200 pixels.  Consequently, pixels with a value
$<0.19~\mathrm{e}^{-}/\mathrm{s}$ (corresponding to a surface
brightness $\mu> 21.02$~mag~arcsec$^{-2}$) were flagged ``low
background''.  Pixel values $\ge0.19~\mathrm{e}^{-}/\mathrm{s}$ and
$<0.25~\mathrm{e}^{-}/\mathrm{s}$
($20.72<\mu<21.02$~mag~arcsec$^{-2}$) were flagged ``intermediate
background'' and pixels with a value
$\ge0.25~\mathrm{e}^{-}/\mathrm{s}$ ($\mu <20.72$~mag~arcsec$^{-2}$) were
flagged ``high background''. These values were chosen because they
resulted in a contour map, shown in Fig.~\ref{fig:Backgroundregions},
in which the high background region clearly follows most of the inner
spiral arms, covering most areas that would be marked ``high
background'' if the selection would take place by eye. The
``intermediate'' region should be considered as a transition region to
clearly separate cluster samples within a low and high background
region.

\subsection{Point spread function}
  \label{subsec:Point spread function}

For our photometry, radius measurements and artificial cluster
experiments we need a filter dependent PSF. Since there are not enough
bright, isolated point sources in the M51 mosaic image to determine
the PSF accurately, the PSF was observationally determined from a
crowded star field on a drizzled image of the Galactic globular
cluster 47~Tuc (NGC~104, \hst\ proposal GO-9281, PI:~J.~E.~Grindlay).
For each filter a separate PSF was determined.

With drizzled data, the size of the PSF depends on the choice of the
drizzle kernel and the accuracy with which the flux from multiple
images is combined. We note that the image of 47~Tuc was drizzled in a
slightly different way than the M51 image. The 47~Tuc images were
drizzled using a square kernel with a size of one pixel
(i.e. $\mathrm{pixfrac} = 1.0$ \citep{fruchter02}), while a Gaussian
kernel with $\mathrm{pixfrac} = 0.9$ was used for M51. Therefore, we
expect some differences between the PSFs, with the M51 PSF possibly being
larger than the 47~Tuc PSF. This could lead to an overestimation of
the measured radii. However, tests by \citet{larsen04b} have shown
that the lower limit down to which \ishape\ can detect a source as
being resolved is $\sim$10\% of the FWHM of the PSF.  At the distance
of M51 and with a FWHM of the ACS PSF of $\sim$2.1 pixels, this
corresponds to an effective radius (\reff) of $\sim$0.5~pc.  We indeed
find a very strong peak in the \reff\ distribution of all the measured
sources below 0.5~pc, consistent with the majority of the sources
(faint stars) being fit as point sources.

 This peak of point sources implies that the measured radii are
\emph{not} biased towards larger values.  We therefore conclude that
the empirical PSF we use, although drizzled in a slightly different
way than the image of M51, is \emph{not} too small.  It shows that
other effects on the PSF, like combining the flux of multiple separate
images, are more important than the differences in the drizzle
parameters.  In \S~\ref{sec:A comparison between ACS and WFPC2} we
also show that there are no large systematic offsets between our
measured radii and the radii of some clusters that were independently
measured by B05 using \wfpc\ data.  We can therefore use the empirical
PSF from the image of 47~Tuc and we will use as a lower-limit for the
measured \reff\ a value of 0.5~pc.

\subsection{Photometry}
  \label{subsec:Photometry}

We performed aperture photometry on all the sources in the source list
using the IRAF\footnote{The Image Reduction and Analysis Facility
(IRAF) is distributed by the National Optical Astronomy Observatories,
which are operated by the Association of Universities for Research in
Astronomy, Inc., under cooperative agreement with the National Science
Foundation.}/DAOPHOT package.  We used a 5 pixel aperture radius and a
background annulus with an inner radius of 10 pixels and a width of 3
pixels.

The aperture correction ($\mathrm{AC}_{0.5\arcsec}$) for resolved
sources from the 5 pixel aperture to 10 pixels ($=0.5\arcsec$) depends
on the size of the source.  Larger sources will have more flux outside
the measuring aperture, and therefore need a larger (i.e. more
negative) aperture correction.  We measured the aperture corrections
on artificial sources with different effective (i.e. projected
half-light) radii between $\sim1$~pc and $\sim5$~pc, generated by the
BAOlab package \citep{larsen99,larsen04b}.  For these artificial
sources we used Moffat profiles \citep{moffat69} with a power-law
index of $-1.5$, which we convolved with the filter dependent PSF of
the \hst/\acs.  The aperture correction was then measured by comparing
the photometry between a 5 and 10 pixel aperture.

\begin{figure}
\centering
 \includegraphics[width=85mm]{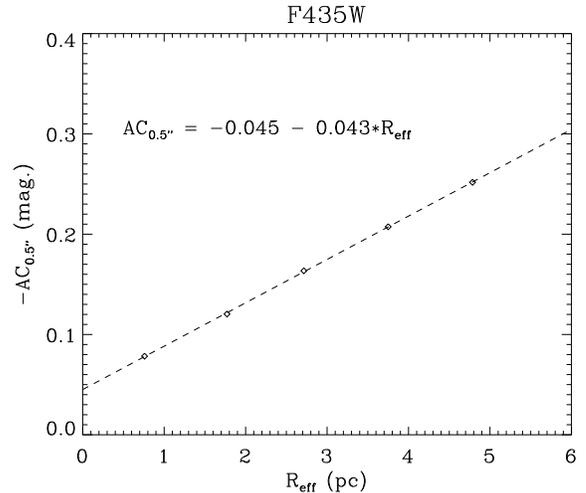}
\caption{The aperture correction ($\mathrm{AC}_{0.5\arcsec}$) for the
flux between a 5 and 10 pixel ($=0.5\arcsec$) aperture in \bband,
versus the effective radius \reff\ of analytical sources at the
distance of M51 (diamonds). The dashed line is a linear fit to the
data of the form of Eq.~\ref{eq:AC}.}
\label{fig:AC}
\end{figure}

The measured aperture corrections in all the three filters (\bband,
\vband\ and \iband) show a clear linear relation with the measured
size of the analytical cluster.  Fig.~\ref{fig:AC} shows the results
for \bband.  The relations between $\mathrm{AC}_{0.5\arcsec}$ and
measured effective radius (\reff) for the different filters can be
written as:
\begin{eqnarray} \label{eq:AC}
 \bband: \frac{\mathrm{AC}_{0.5\arcsec}}{\mathrm{mag}} & = & -0.045-0.043\cdot\left(\frac{\reff}{\mathrm{pc}}\right) \nonumber\\
 \vband: \frac{\mathrm{AC}_{0.5\arcsec}}{\mathrm{mag}} & = & -0.045-0.042\cdot\left(\frac{\reff}{\mathrm{pc}}\right) \\
 \iband: \frac{\mathrm{AC}_{0.5\arcsec}}{\mathrm{mag}} & = & -0.057-0.043\cdot\left(\frac{\reff}{\mathrm{pc}}\right). \nonumber
\end{eqnarray}
For a source which has a typical measured size of 3~pc (see
\S~\ref{sec:Radius distribution}), this corresponds to an aperture
correction of $-0.17$, $-0.17$ and $-0.19$ mag in \bband, \vband\ and
\iband, respectively.

We could use Eq.~\ref{eq:AC} to apply a size-dependent aperture
correction to every source based on their radius measurement.
However, this could introduce new and unknown systematic uncertainties
due to the limitations of the radius measurements.  We therefore
decided to use a fixed aperture correction, determined for a 3~pc
source.  We are aware that this introduces uncertainties in the flux
as a function of the sizes of the sources \citep{anders06}.  We will
underestimate the flux with $\sim$0.3 mag~for a 9~pc source and
overestimate the flux with $\sim$0.1 mag~for a 0.5~pc source (also see
\S~\ref{sec:Artificial cluster experiments}).  However, it is
preferred to deal with these systematic uncertainties of known origin
rather than introducing new uncertainties due to the less well
determined uncertainties of the sizes. Moreover, since the
coefficients in Eq.~\ref{eq:AC} are very similar for the different
filters, uncertainties in the aperture corrections practically cancel
out when we consider colours instead of fluxes. Nevertheless, when
necessary we will mention how our results change if we use the
size-dependent aperture corrections.

The filter dependent aperture corrections from $0.5\arcsec$ to
infinity (the ``infinity corrections'') were taken from Table 5 of
\citet{sirianni05}, which were determined for point sources.  In
principle, with these infinity corrections for point sources we are
slightly underestimating the infinity corrections for resolved
sources.  However, in \S~\ref{sec:Artificial cluster experiments} we
will show, using artificial cluster experiments, that with the
infinity corrections for point sources we are not introducing
systematic offsets in the photometry for 3~pc sources.

A final correction for Galactic foreground extinction of $E(B-V) =
0.038$ in the direction of M51 was applied, according to Appendix B of
\citet{Schlegel98}.  This corresponds to an additional correction in
\bband, \vband\ and \iband\ of $-0.152$, $-0.117$ and $-0.068$ mag,
respectively.

We did not apply any CTE corrections, since on the drizzled mosaic
image the exact location of every source on the CCD is not easily
retrieved, nor did we do photometry on the single
(i.e. ``un-stacked'') exposures, both of which are necessary to
calculate the CTE corrections.  We have estimated the CTE corrections
to be of the order of $-0.02$ mag and therefore ignoring them does not
lead to large systematic effects.  We also did not apply a correction
for the impact of the \emph{red halo} \citep{sirianni05} on our
\iband\ photometry, since clusters in the spiral disc of M51 are
mainly blue objects and the red halo effect is most significant for
very red objects observed in the \emph{F850LP} filter.  Using Tables 6
and 8 from \citet{sirianni05} we estimate that the error in the
infinity correction for our \iband\ photometry would be of the order
of $\sim0.01$~mag, if the clusters would be red objects.  This shows
that the red halo effect has no significant effect on our photometry.

\section{Radius measurements}
  \label{sec:Radius measurements}

We exploit the resolution of the \acs\ camera by measuring the
effective radii of all the $75\,436$ detected sources in \bband,
\vband\ and \iband.  These radii can than be used to distinguish
\emph{slightly resolved} stellar cluster candidates from stars
(\S~\ref{sec:Selection of the sample}) and to study the size
distribution of a large sample of stellar cluster candidates
(\S~\ref{sec:Radius distribution}).  With ``slightly resolved'' we
mean that the cluster candidates have an intrinsic size which is
comparable to or smaller than the FWHM of the PSF.
 
For the radius measurements we used the \ishape\ routine, which is
part of the BAOlab package \citep{larsen99,larsen04b}.  \ishape\
convolves analytic profiles for the surface brightness distribution of
a cluster with different effective radii with the PSF and then fits
these to each source in the data.  The best fitting \reff\ is then
determined by minimizing the $\chi^2$ in an iterative process.  For
the analytic profiles we used the same ones as for the measured
aperture corrections described in \S~\ref{subsec:Photometry}, namely
Moffat profiles with a power-law index of $-1.5$ (i.e. a Moffat 15
profile).  These profiles were found to be the best-fitting profiles
to young stellar clusters in the LMC \citep{elson87}.

Because the M51 \acs\ data was drizzled, the cores of the surface
brightness profiles of the young clusters in M51 could have been
slightly changed.  We did not quantify this effect, but instead stick
to the Moffat 15 profiles, since the outer regions of the profiles,
which in the case of Moffat 15 profiles approximate power laws, are not
expected to change.

The average surface brightness profile of Galactic globular clusters
(GCs) is a King 30 profile \citep{king62, harris96}.  Tests by
\citet{larsen99} have shown that when clusters that follow a King
profile are measured using a Moffat 15 profile, the effective radius
is reproduced quite well.  Therefore, even in the case where the true
profiles of stellar clusters in M51 are not perfect Moffat 15
profiles, the use of this profile will likely not lead to large
systematic errors.

The radii of the sources were measured using the flux within a 5 pixel
radius around the centre of the source (i.e. using an \ishape\ fitting
radius of 5 pixels).  To avoid neighbouring sources to affect the
radius measurements, we rejected all sources which have a neighbour in
the original source list within 5 pixels (see \S~\ref{sec:Selection of
the sample} for a full description of the applied selection criteria).

Since stellar cluster profiles are almost never perfectly spherically
symmetric, we fitted them with elliptical profiles.  We transformed
the measured FWHM (in pixels) along the major axis into an effective
radius according to the formulae from the \ishape\ user's guide:
\begin{equation} \label{eq:effective radius}
\reff \approx 0.5(1+\mathrm{ratio})\times1.13\times2.036\cdot \mathrm{FWHM},
\end{equation}
which gives \reff\ in parsecs.  The factor 2.036 accounts for the size
of a pixel in parsecs at the distance of M51, the factor 1.13 is the
conversion factor from FWHM to \reff\ for a Moffat 15 profile and the
term $0.5(1+\mathrm{ratio})$ is a correction for the elliptical
profile (using the derived aspect ratio).  Because of the correction
for ellipticity, Eq.~\ref{eq:effective radius} gives the radius of a
circular aperture containing half the total light of an elliptical
profile.  This way we have a single value for the effective radius of
elliptical sources and we still preserve information about the aspect
ratios and position angles of the sources for future studies.

\section{Artificial cluster experiments}
  \label{sec:Artificial cluster experiments}

To test how our selection of stellar cluster candidates
(\S~\ref{sec:Selection of the sample}) depends on the brightness and
size of the cluster and the background region, and how accurate our
radius measurements are, we performed tests with artificial clusters
for all three filters, different background regions and different
cluster sizes.  The results of these tests will be used in
\S~\ref{sec:Selection of the sample} to select a sample of stellar
cluster candidates with accurate radii.

First we created artificial clusters using the MKCMPPSF and MKSYNTH
tasks, which are part of the BAOlab package \citep{larsen99}. For
every filter we convolved the PSF with a Moffat profile with a
power-law index of $-1.5$ and effective radii between 1 and 9~pc, in
steps of 2~pc.  These artificial clusters were then scaled to a range
of magnitudes between 18 and 26 mag with steps of 0.5 mag.  For every
magnitude 100 clusters were added at random locations to every
background region on the mosaic image by combining MKSYNTH with the
\texttt{imarith} task in IRAF.  We made sure that the minimum distance
between these random locations was at least 30 pixels, in order not to
introduce artificial crowding effects.

We then performed the cluster selection on these sources in a similar
way as with the normal data.  We performed these tests for every
filter individually, since taking into account the matching of every
source in three filters, as we did with the normal data, would imply
creating artificial clusters with a range of colours (i.e. ages) for
every magnitude, drastically increasing the computing time.  However,
by comparing analytical spectral energy distributions (SEDs) from
GALEV simple stellar population (SSP) models \citep{schulz02} to the
measured detection limits, we tested that for cluster ages up to
$\sim$1~Gyr, the \iband\ filter is always the most limiting filter for
detecting clusters.  We therefore do not expect that using the results
of these tests for individual filters is introducing large
age-dependent biases in the derived detection limits.

\begin{figure}[t]
\centering
 \includegraphics[width=85mm]{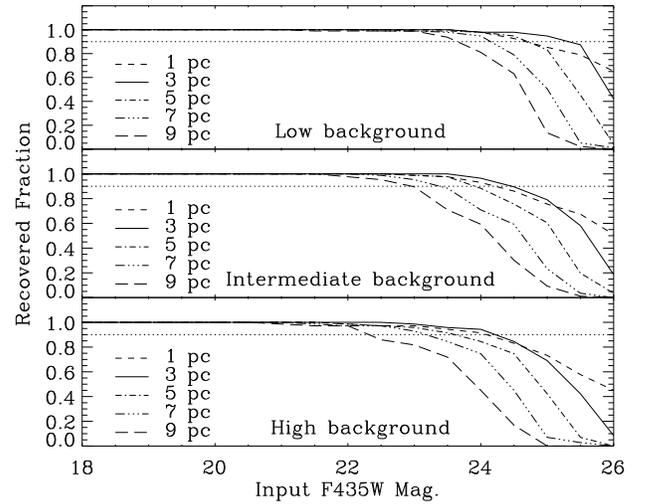}
\caption{The recovered fraction of artificial clusters as a function
of magnitude and \reff\ for the three background regions on the
\bband\ image (i.e.\ ``completeness profiles''). The horizontal dotted
line indicates where 90\% completeness is reached. The magnitudes at
which the completeness profiles reach this 90\% completeness limit are
summarized in Fig.~\ref{fig:limits}.}
\label{fig:completeness_regions}
\end{figure}

The artificial clusters were recovered by running SExtractor, followed
by photometry in all three filters and by performing size measurements
in \bband\ and \vband.  For \bband\ and \vband, we considered a
cluster to be recovered when we found a resolved source (i.e. with a
$\reff$ larger than our adopted lower limit of 0.5~pc (see
\S~\ref{sec:Selection of the sample}) and a $\chi^2$ which is lower
than the $\chi^2$ using only a PSF) within 1 pixel from the input
coordinate and with a distance to the nearest neighbouring source of
$>$5~pixels.  For \iband\ we considered a cluster to be recovered when
we found a source within 1 pixel from the input coordinate and with a
distance to the nearest neighbouring source of $>$5~pixels.
The recovered fraction as a function of \bband\ magnitude for the
different background regions and input radii is plotted in
Fig.~\ref{fig:completeness_regions}.  For the \vband\ and \iband\
passbands the results are similar, except that the turn-off of the
completeness curves happens at brighter magnitudes due to a lower S/N
ratio of the \vband\ and \iband\ data (see below).

The recovered fraction shown in Fig.~\ref{fig:completeness_regions} is
scaled to the number of clusters recovered at magnitude 18.  We note,
however, that a certain fraction of even the brightest artificial
clusters is not recovered due to their vicinity within 5 pixels from a
neighbouring source.  Our completeness tests show that these initial
losses will be $\sim$3, $\sim$13 and $\sim$28\% for the low,
intermediate and high background, respectively.  This shows that due
to crowding effects one can never select a sample which is 100\%
complete.  Depending on the selection criteria and background region,
one will lose up to $\sim$28\% of the initial sample present in the
data.  This number will be even higher when one considers that young
clusters are not randomly distributed across the spiral disc, but will
mostly be clustered themselves in high background regions.

\begin{figure}[t]
\centering
 \includegraphics[width=85mm]{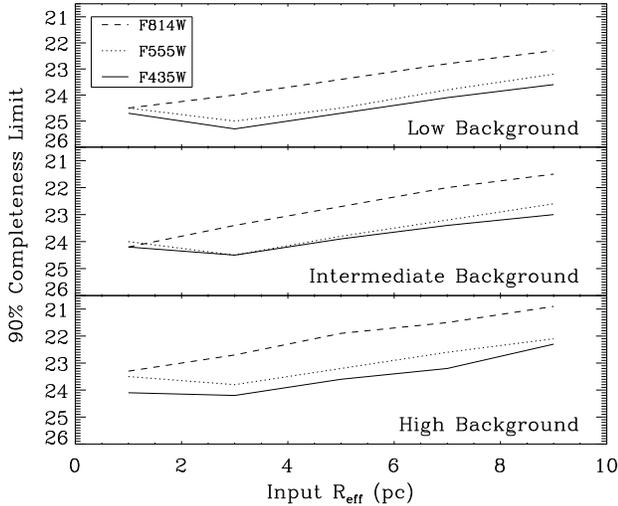}
\caption{The 90\% completeness limits versus effective radius (\reff),
determined for \bband, \vband\ and \iband\ for the three background
regions.}
\label{fig:limits}
\end{figure}

Fig.~\ref{fig:completeness_regions} shows that larger clusters are
harder to recover than smaller clusters with the same brightness.
This is expected, since larger clusters have a lower surface
brightness, which makes them easier to blend into the background.  We
define the magnitude at which 90\% of the artificial cluster was
recovered to be the 90\% completeness limit.  The values we found this
way for the different filters, cluster sizes and background regions
are plotted in Fig.~\ref{fig:limits}.

\begin{figure}[t]
\centering
 \includegraphics[width=85mm]{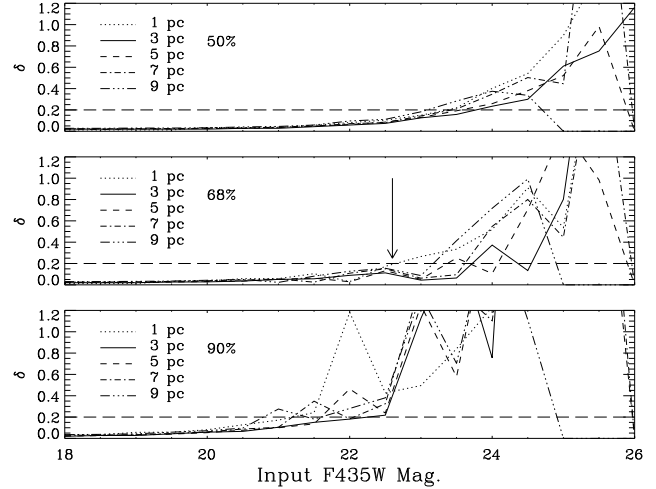}
\caption{The median (top), 68th percentile (middle) and 90th
percentile (bottom) of $\delta$ (see Eq.~\ref{eq:delta}) versus
\bband\ magnitude for the artificial clusters with different radii in
the high background region. Below the horizontal line the uncertainty
of the radii measurements is smaller than 20\%, which is achieved for
more than 68\% of the clusters up to 9 pc brighter than $\bband =
22.6$~mag (indicated by the arrow).}
\label{fig:completeness_sizedifferences}
\end{figure}

For our current study, where we look at the radius distribution of
stellar clusters, it is not only important to detect clusters by
measuring their radii, but the measured radii also have to be
accurate.  To test the limitations of our radius measurements, we
looked at how the difference between input and measured radius of the
artificial clusters depends on magnitude, input size and background
region.  In Fig.~\ref{fig:completeness_sizedifferences} we plot the
50th (i.e.\ the median), 68th and 90th percentile of $\delta$ versus
the magnitude in \bband, where we define $\delta$ as the relative
difference between input and measured radius:
\begin{equation} \label{eq:delta}
\delta \equiv
\frac{|R_{\mathrm{eff,in}}-R_{\mathrm{eff,out}}|}{R_{\mathrm{eff,in}}}.
\end{equation}
We note that the \emph{p}th percentile is the value such that
\emph{p}\% of the observations ($\delta$) is less than this value.
Fig.~\ref{fig:completeness_sizedifferences} shows for example, that
when we select cluster candidates with $\bband<22.6$ mag, $\sim$68\%
of the sample will have a radius uncertainty smaller than 20\%.  The
figure also indicates that the radius measurements are the most
limiting factor in the detection of stellar cluster candidates: at the
90\% completeness limit for a 3~pc source ($\bband = 24.2$~mag), about
50\% of the recovered sources are likely to have inaccuracies in their
radii \emph{larger} than 25--40\%.  Therefore, in
\S~\ref{sec:Selection of the sample} we will select magnitude limits
in \bband\ and \vband\ \emph{brighter} than the 90\% completeness
limits of these passbands.  Since we will not use the radius
measurements in the \iband\ passband, in this passband the 90\%
completeness limit will be used in the selection of the sample.

\begin{figure}[t]
\centering
 \includegraphics[width=85mm]{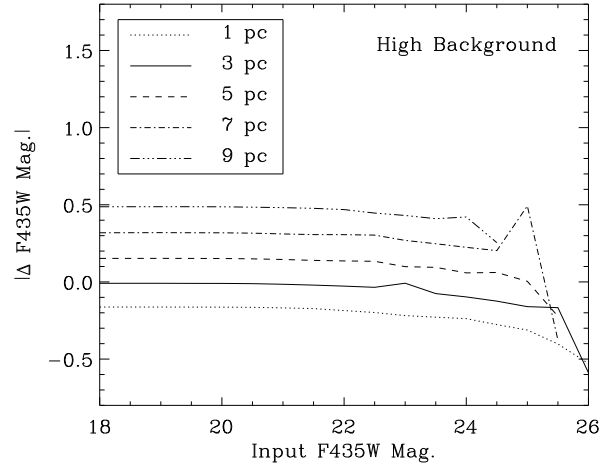}
\caption{The mean difference between the measured magnitude and the
input magnitude of the recovered artificial clusters versus \bband\
magnitude for different sizes in the high background region. The
truncated line for the largest and faintest clusters means none of
these artificial clusters were recovered.}
\label{fig:Delta mag}
\end{figure}

We also used the results of our artificial cluster experiments to test
the robustness of our photometry and the accuracy of the applied
aperture corrections described in \S~\ref{subsec:Photometry}.  In
Fig.~\ref{fig:Delta mag} we show the mean difference between the
measured magnitude and the input magnitude ($\Delta$~mag) versus
\bband\ magnitude for different sizes in the high background region.
We applied the constant aperture correction for a 3~pc source
according to Eq.~\ref{eq:AC} and the infinity correction for point
sources from \citet{sirianni05} that we described in
\S~\ref{subsec:Photometry}.  First of all, Fig.~\ref{fig:Delta mag}
shows that the applied aperture correction is very accurate, since the
photometry of 3~pc clusters is almost perfectly reproduced to
$\sim$22.5~mag.  This shows that applying the point source infinity
correction to 3~pc sources does not introduce systematic offsets in
the photometry.  Second, the range covered in $\Delta$~mag shows that
when there is no information about the radius of the cluster, the
uncertainty in the photometry can be as large as $\sim0.5$~mag for
clusters with radii in the range 1--9~pc.

\section{Selection of the sample}
  \label{sec:Selection of the sample}

We used the radius estimates to distinguish the resolved
clusters from unresolved objects.  In this section we select two
different cluster samples: a ``resolved sample'' with clearly resolved
clusters, and a smaller subset from this sample, a ``radius sample''.
The radius sample satisfies extra criteria that make the radii more
reliable, and will be used to study the radius distribution
(\S~\ref{sec:Radius distribution}) and the correlations between mass,
radius and distance (\S~\ref{sec:Correlations between mass, radius and
distance}).  The resolved sample will be used in another study of the
luminosity function of stellar clusters in M51 \citep[in prep.]{gieles06a,
haas07}.

The cluster selection process is hampered by various factors like an
irregular background (spiral arms, dust lanes), contaminating
background galaxies and crowding effects which causes many sources to
be (partially) blended with neighbouring sources, biasing the radius
measurements.  We tried to automate the selection of stellar cluster
candidates as much as possible, taking into account all these factors.
However, it was unavoidable to subject the automatically selected
sample to visual inspection, to filter out any remaining contaminants
or the stellar cluster candidates of which the radii measurements
could not be trusted.

\subsection{Selection criteria}
  \label{subsec:Selection criteria}

\begin{enumerate}

\item Our first two selection criteria were concerned with the sources
actually being resolved.  As mentioned in \S~\ref{sec:Radius
measurements}, we will use as a lower-limit for the measured \reff\ a
value of 0.5~pc, since below this radius we observe a strong peak of
unresolved sources (most likely bright stars).  Therefore, our first criterion for the selection
of stellar cluster candidates was to select sources with $\reff >
0.5$~pc.  We applied this criterion to the radius measurements in both
\bband\ and \vband.  We did not apply a criterion to the measured
radii in \iband, since that would have restricted our sample too much
due to the lower signal-to-noise ratio in this passband.

\item Not only should a stellar cluster candidate be resolved when we
fit a Moffat15 profile, but using the profile convolved with the PSF should also result in a better fit
than fitting the candidate with the PSF alone. A second criterion was therefore to use the $\chi^{2}$ of the
radius fit using the Moffat15 profile, which should be lower than the $\chi^{2}_{0}$ of a fit
using only the PSF: $ \chi^{2}/\chi_{0}^{2} < 1$.
\end{enumerate}

\begin{figure}[t] 
\centering
\begin{minipage}[center]{85mm}
\includegraphics[width=27mm]{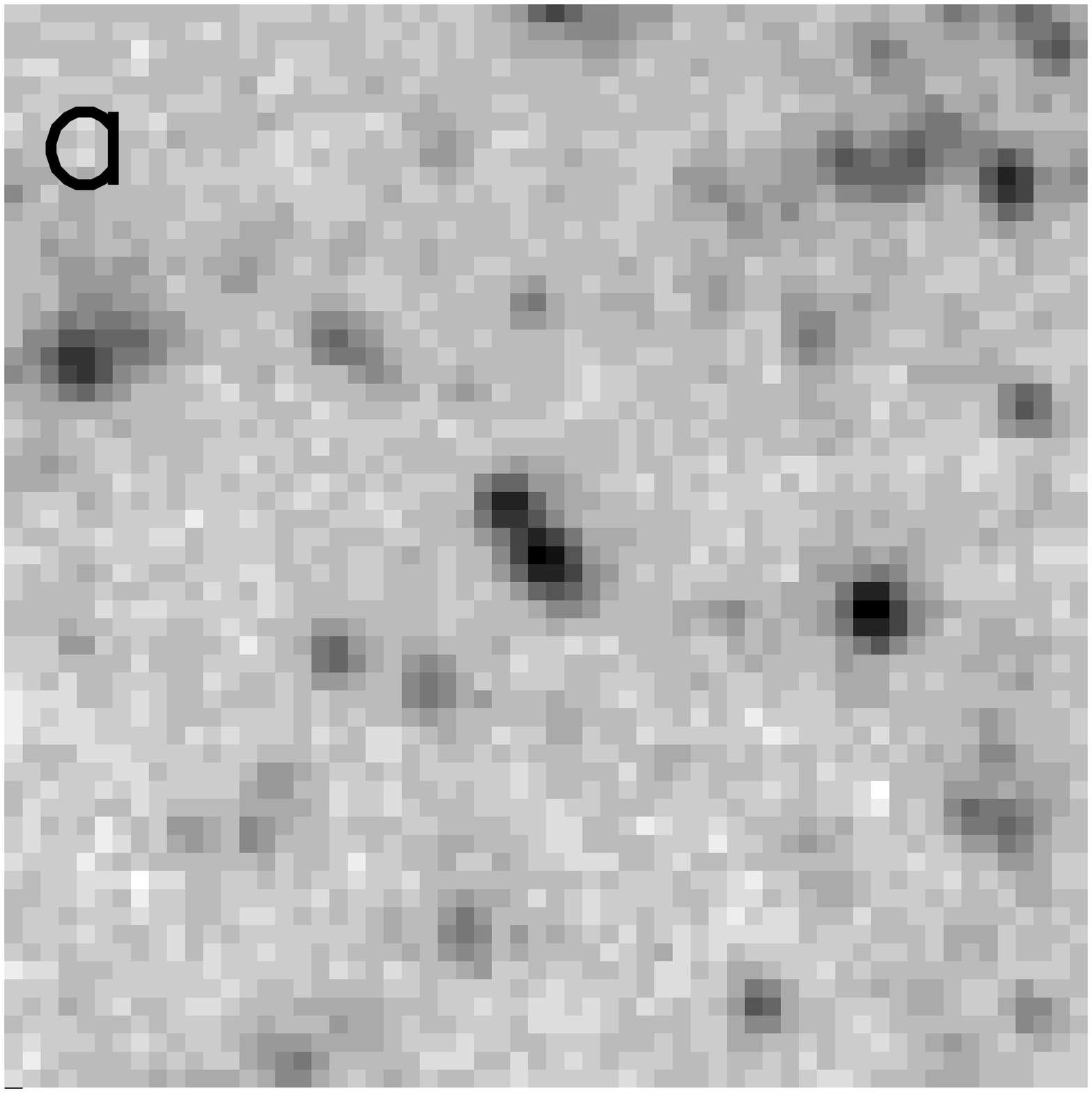}
\includegraphics[width=27mm]{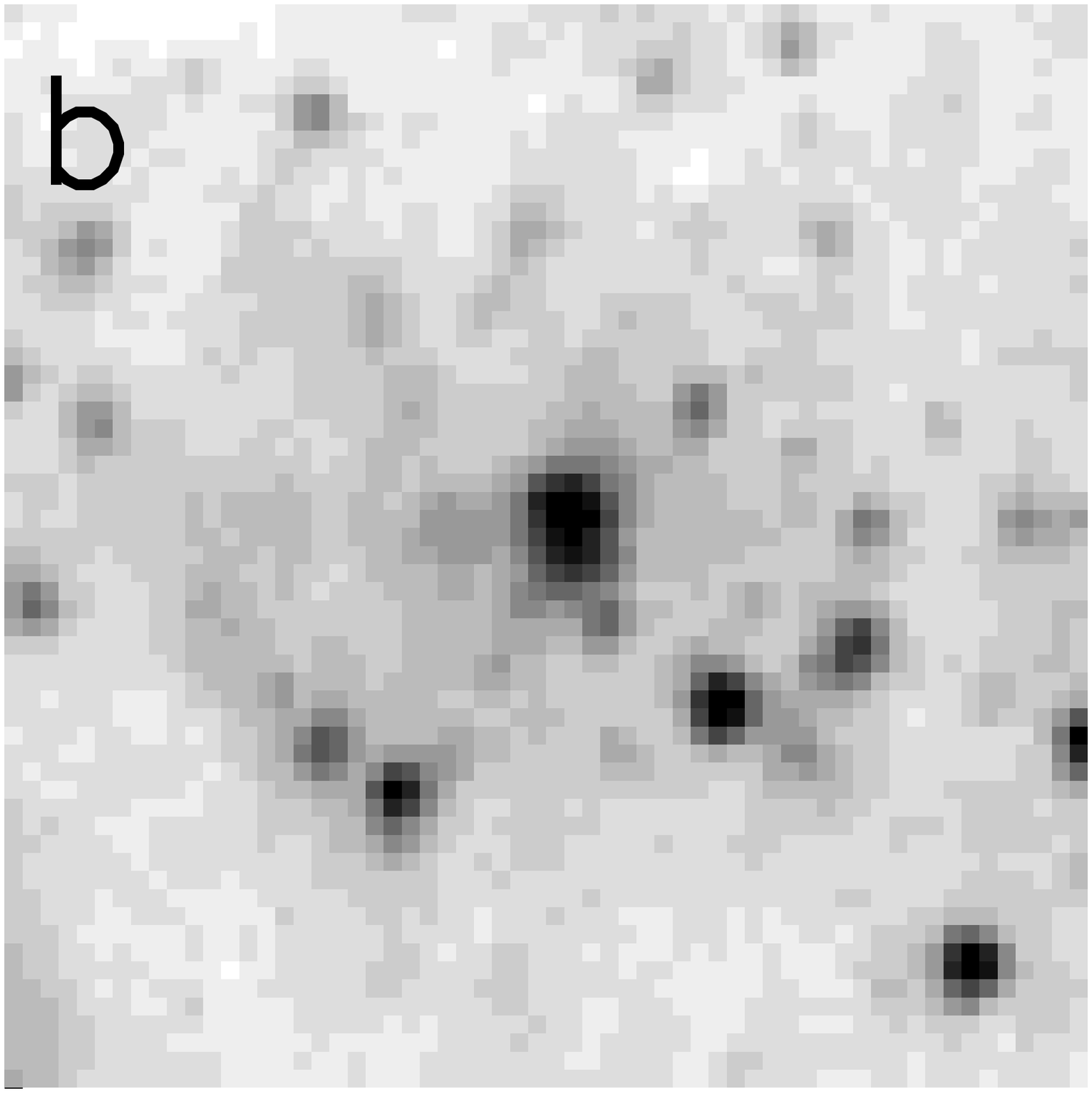}
\includegraphics[width=27mm]{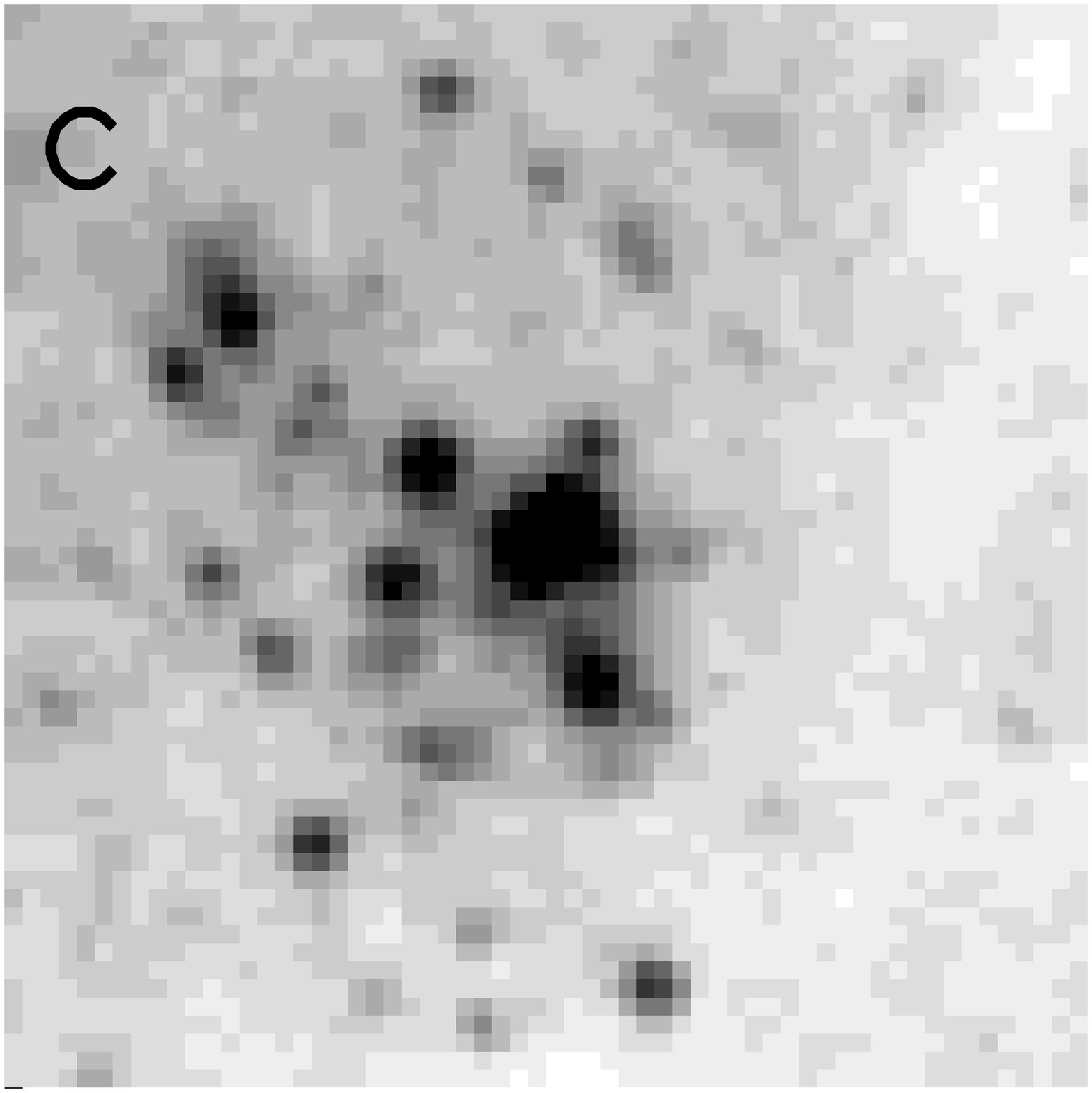}\\
\includegraphics[width=27mm]{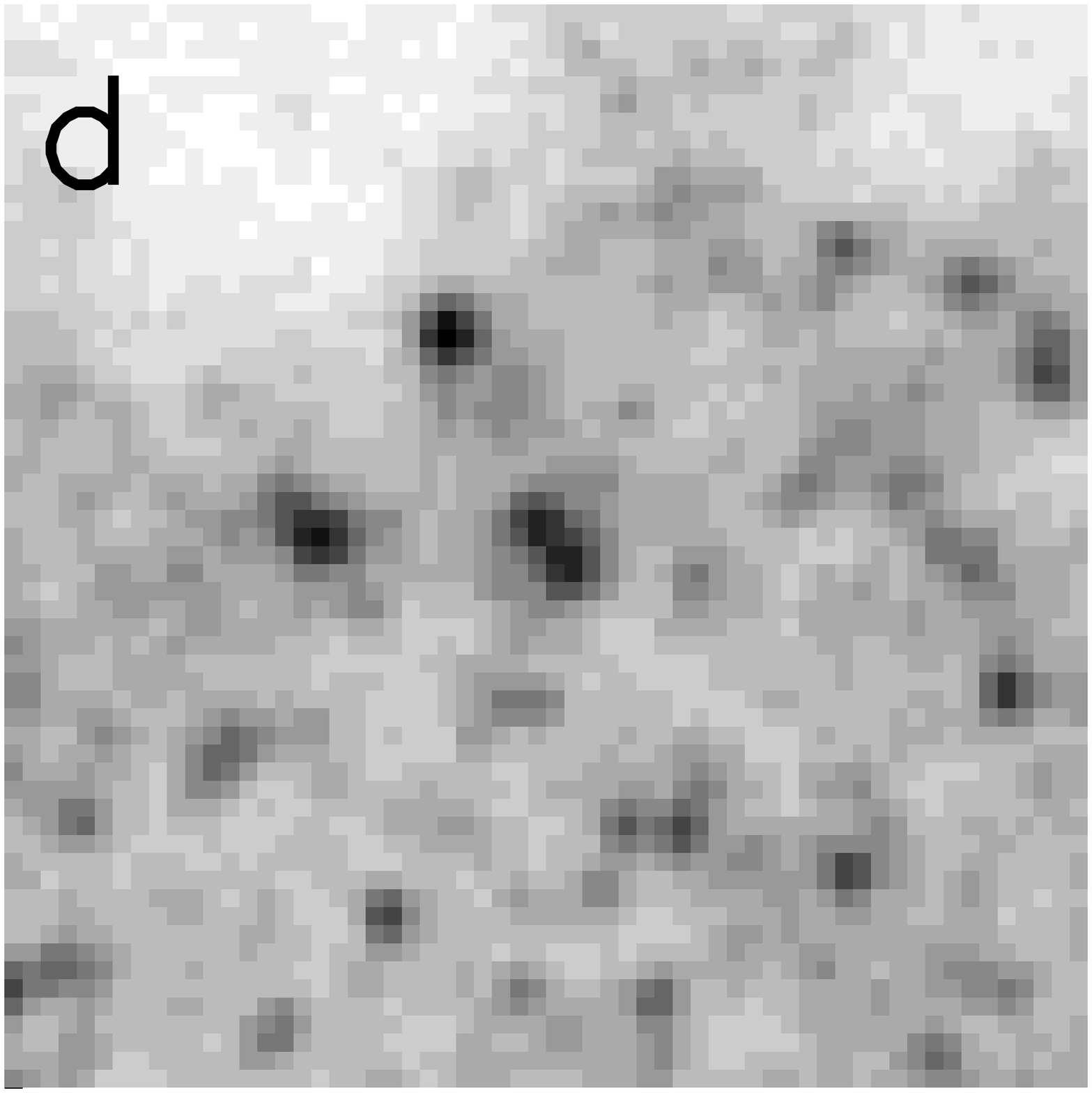}
\includegraphics[width=27mm]{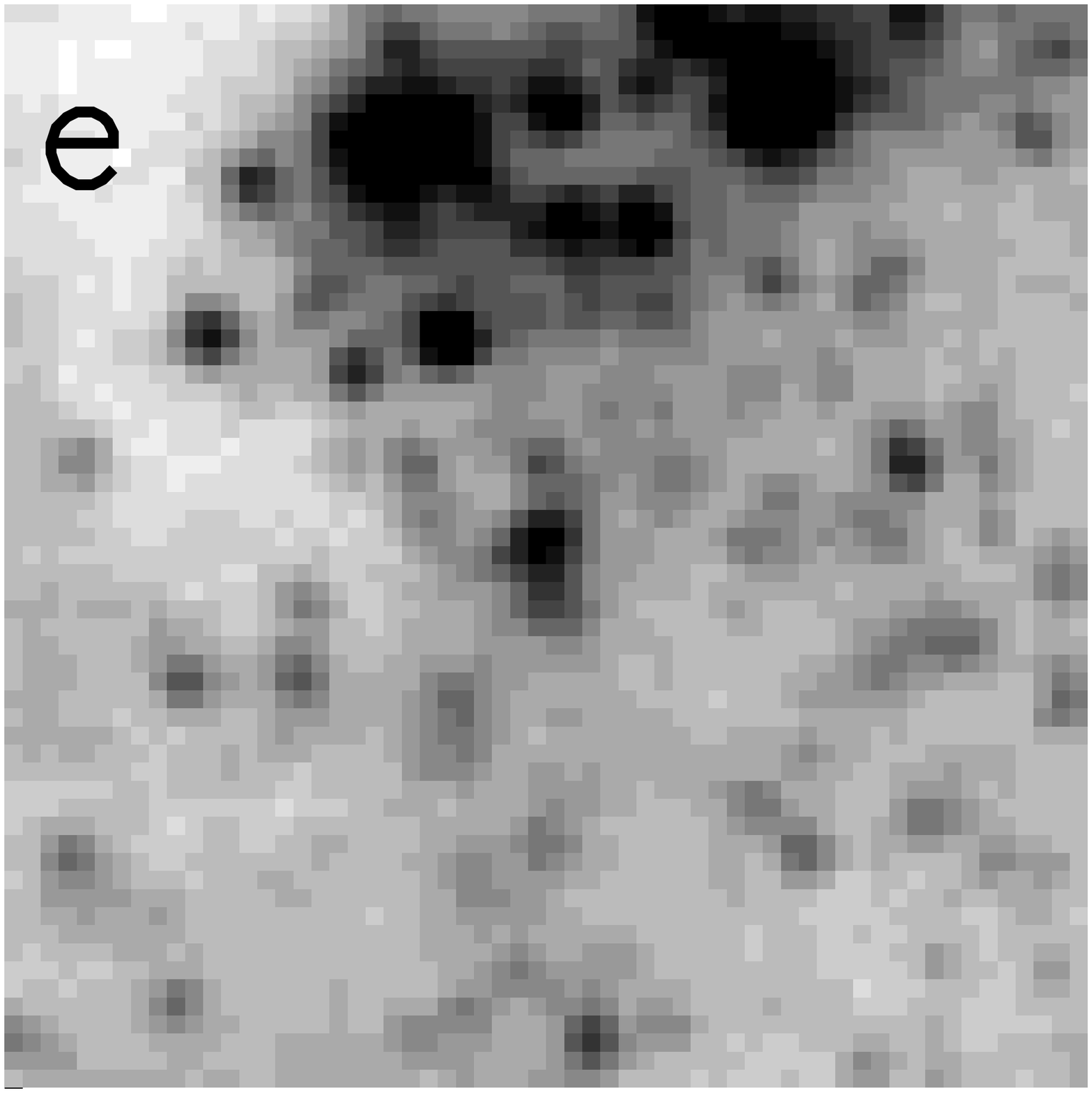}
\includegraphics[width=27mm]{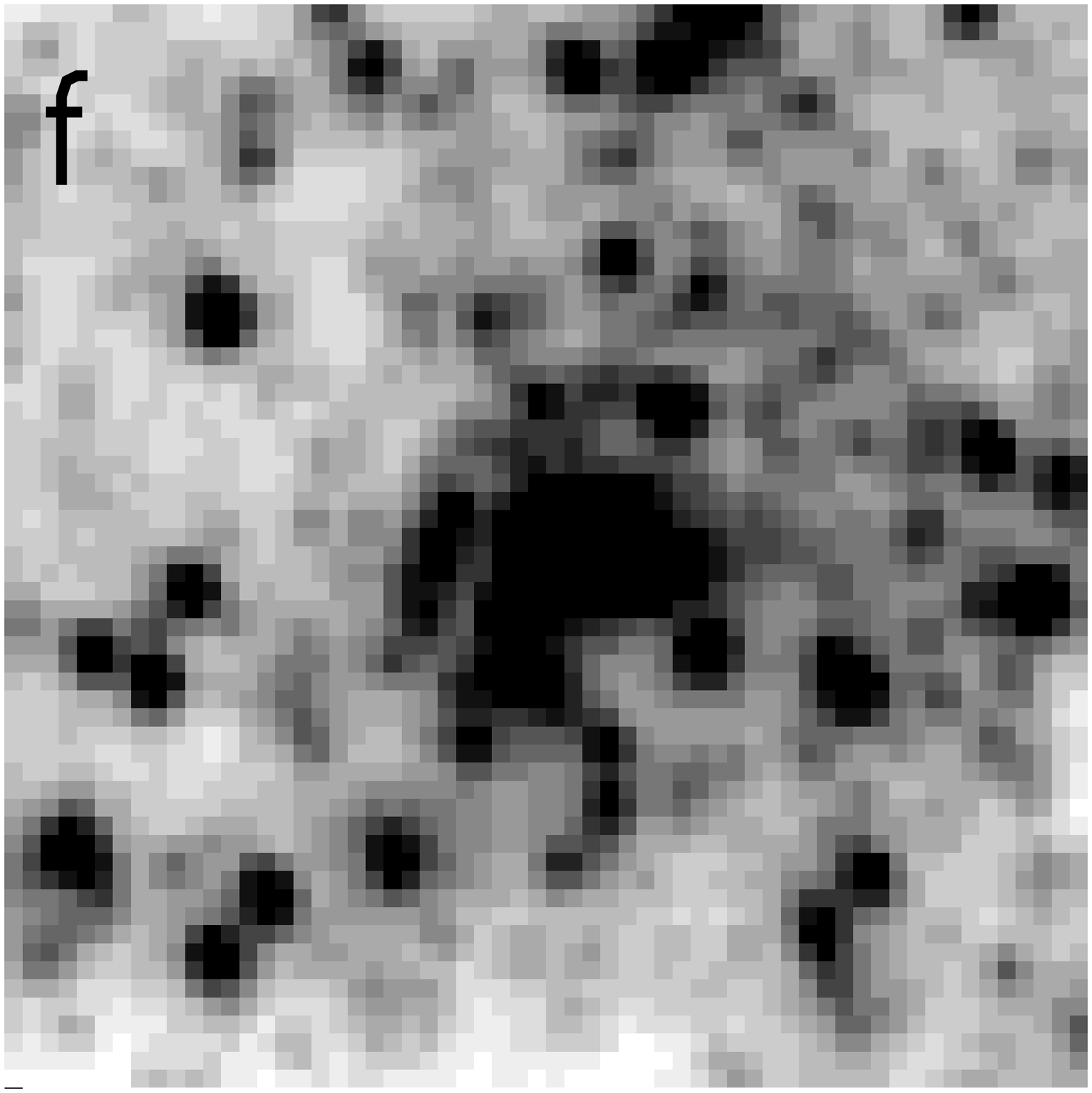}\\
\includegraphics[width=27mm]{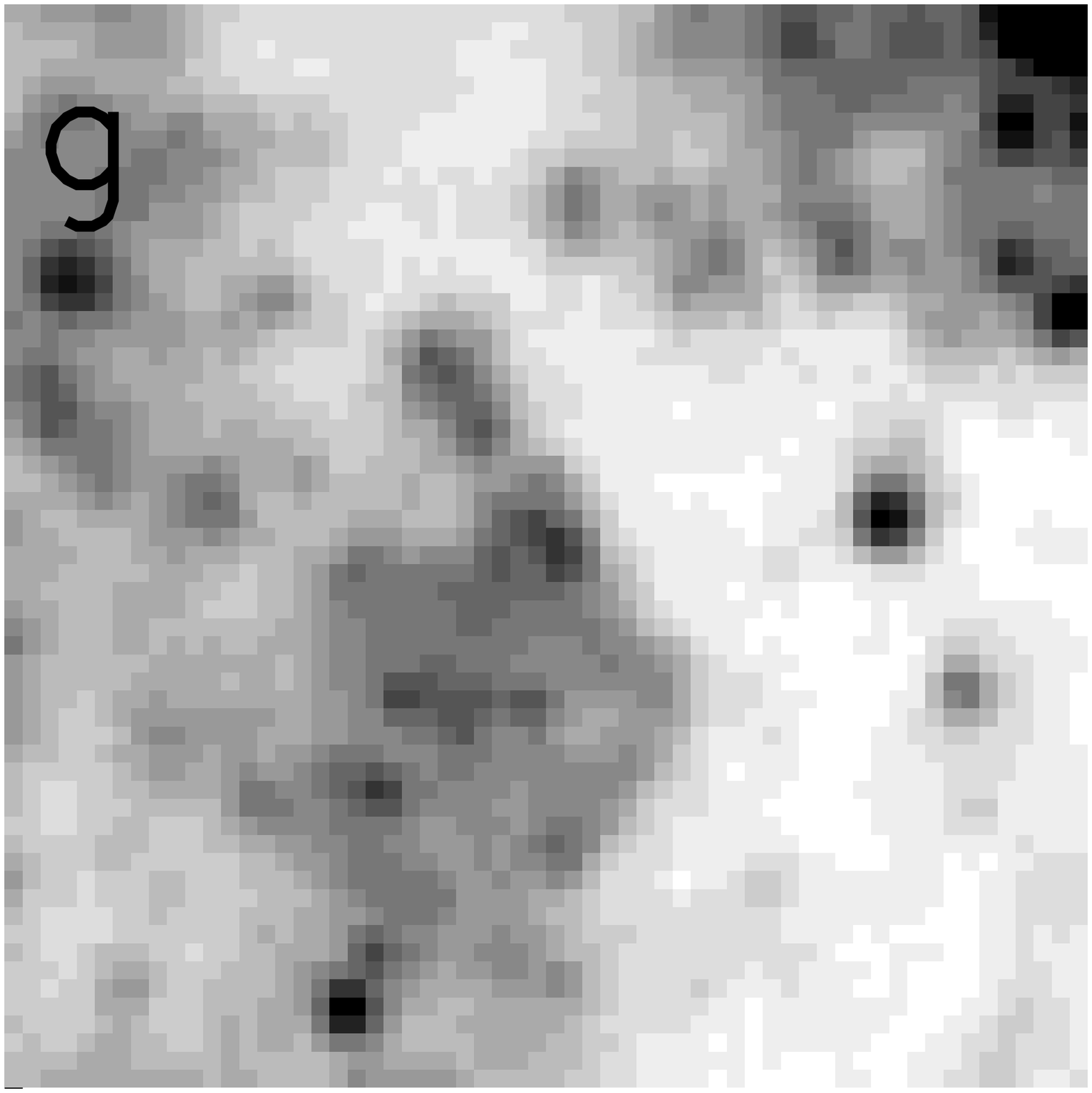}
\includegraphics[width=27mm]{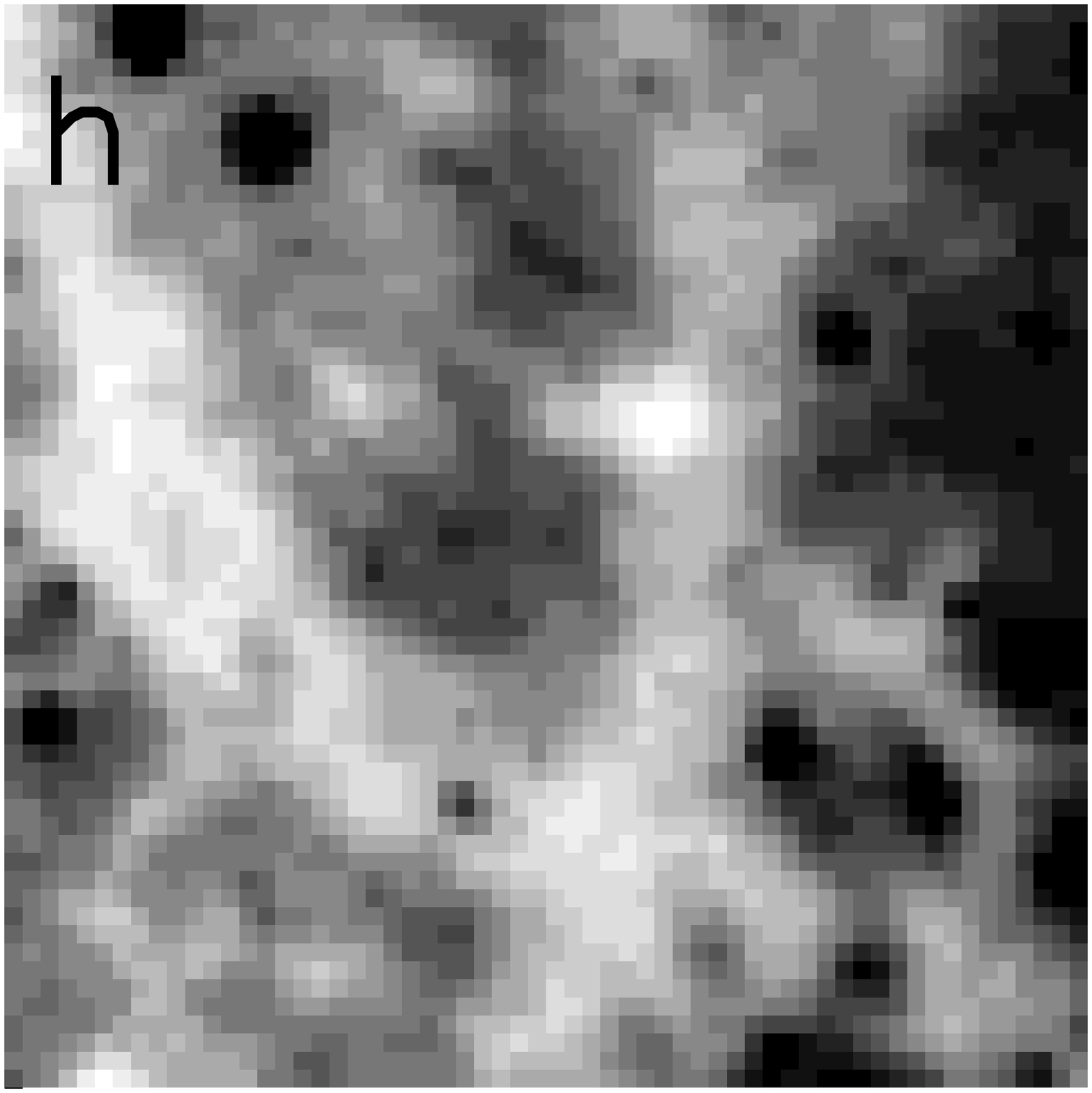}
\includegraphics[width=27mm]{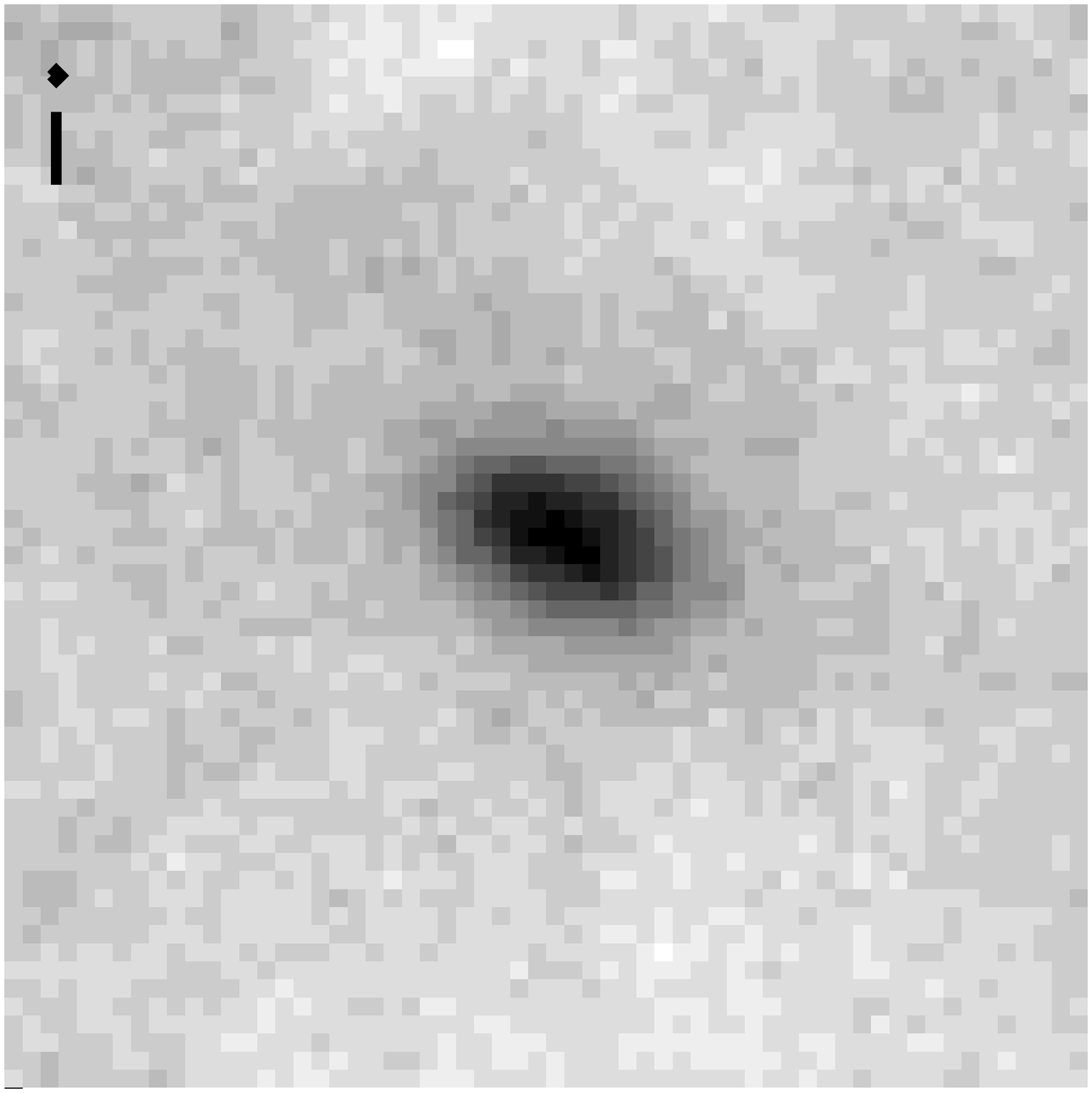}
\caption{A selection of \bband\ images centered on sources that were
rejected from the sample after visual inspection. The size of each
image is 60$\times$60 pixels or 3\arcsec$\times$3\arcsec. Rejection
was based on the following criteria: a.~neighbour too close by,
b.~likely blended, c~\&~d.~clearly blended, e.~too
irregular, f.~crowded, g.~interfering background, h.~radius fit on the background and i.~likely a background galaxy.}
\label{fig:crap}
\end{minipage}
\end{figure}

\begin{figure}[t] 
\centering
\begin{minipage}[center]{85mm} 
\includegraphics[width=27mm]{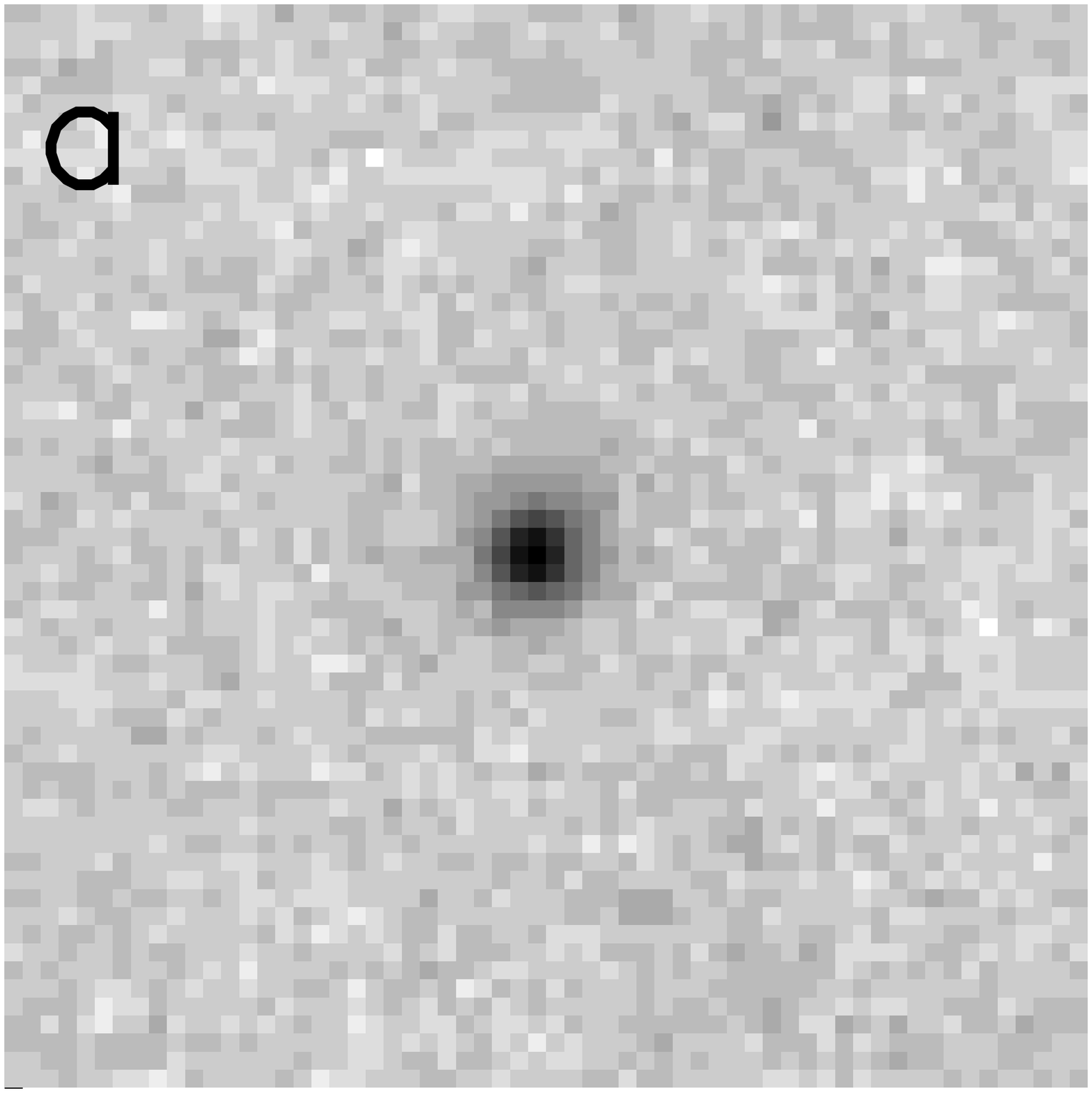}
\includegraphics[width=27mm]{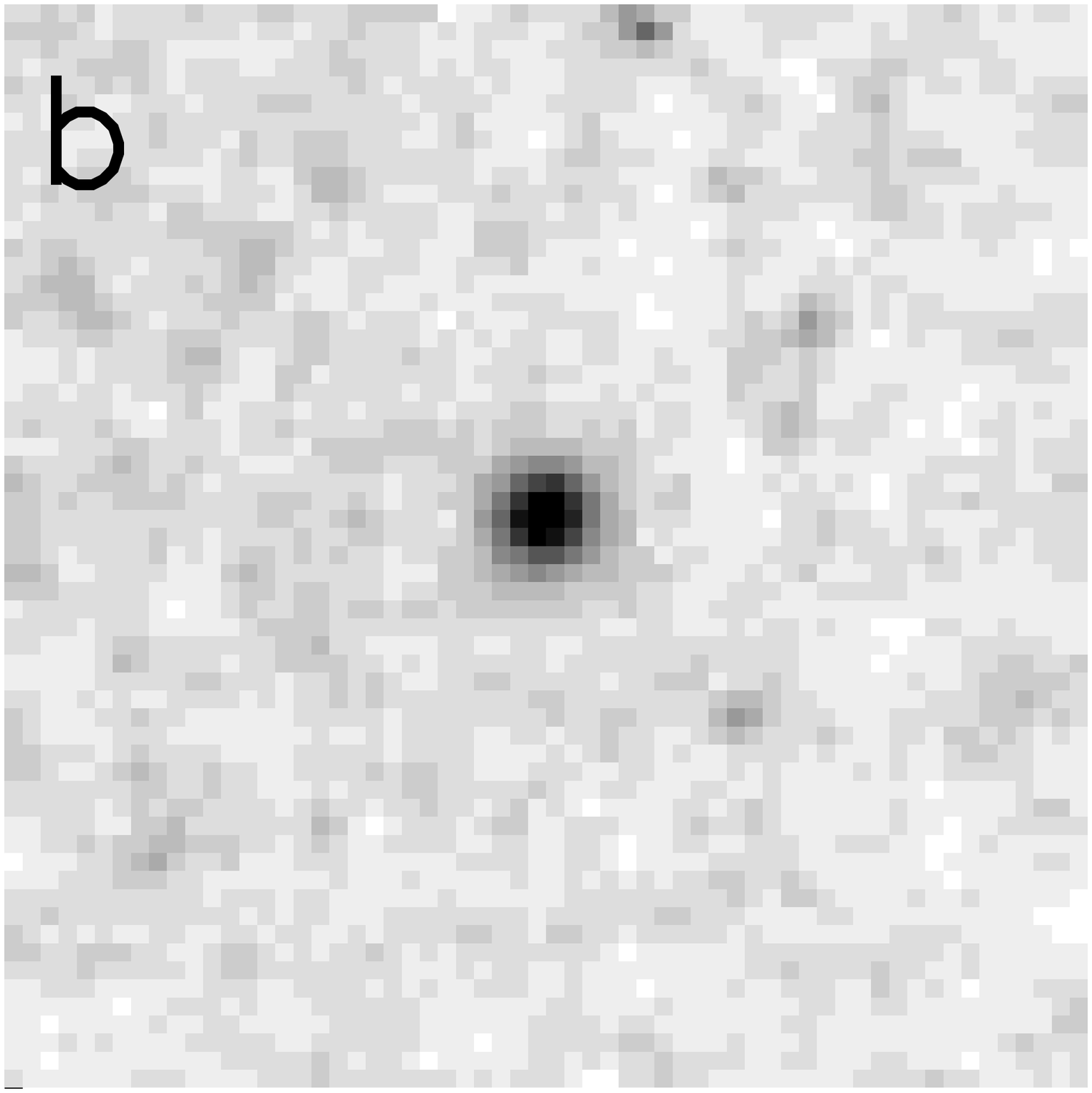}
\includegraphics[width=27mm]{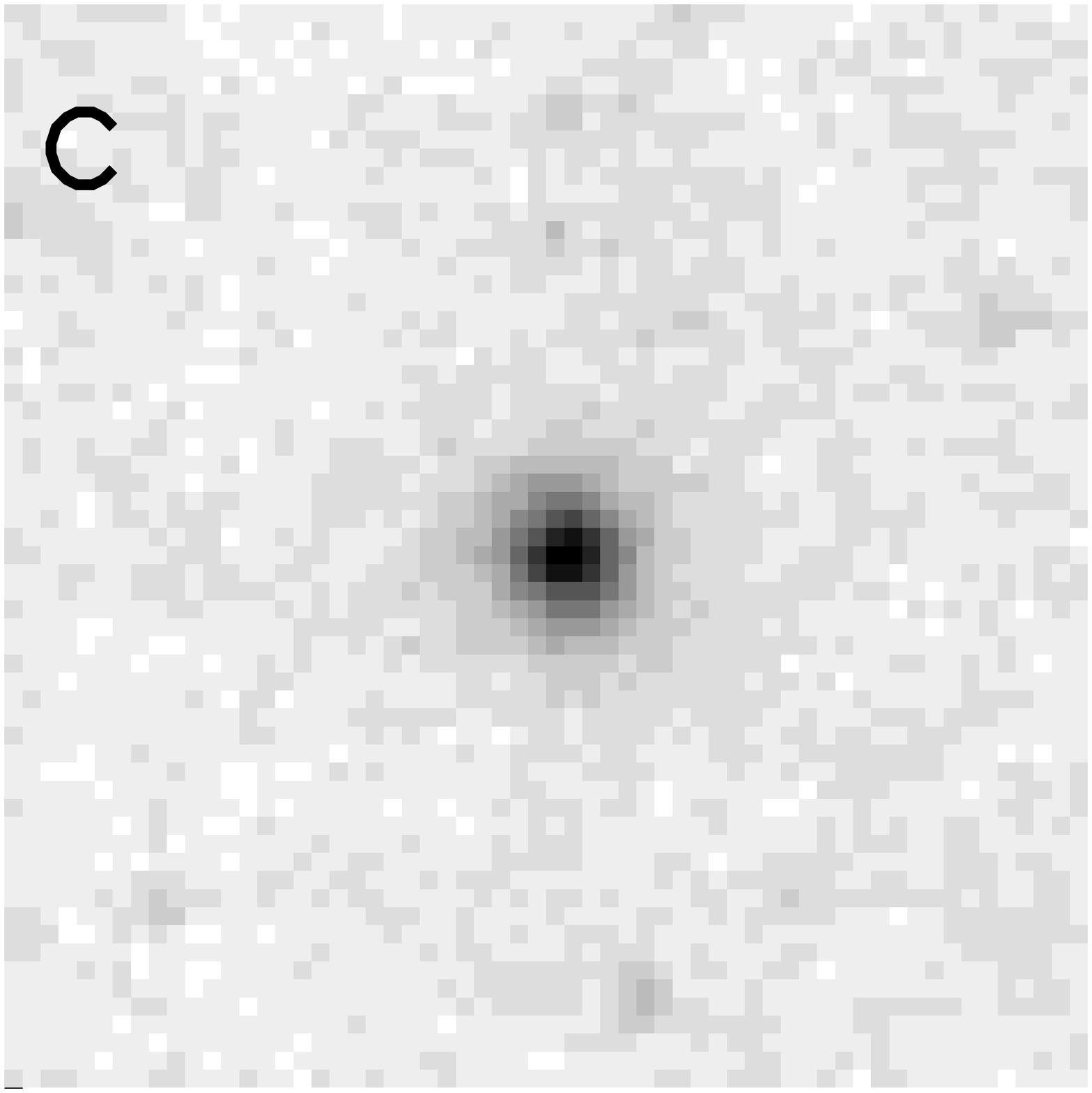}\\
\includegraphics[width=27mm]{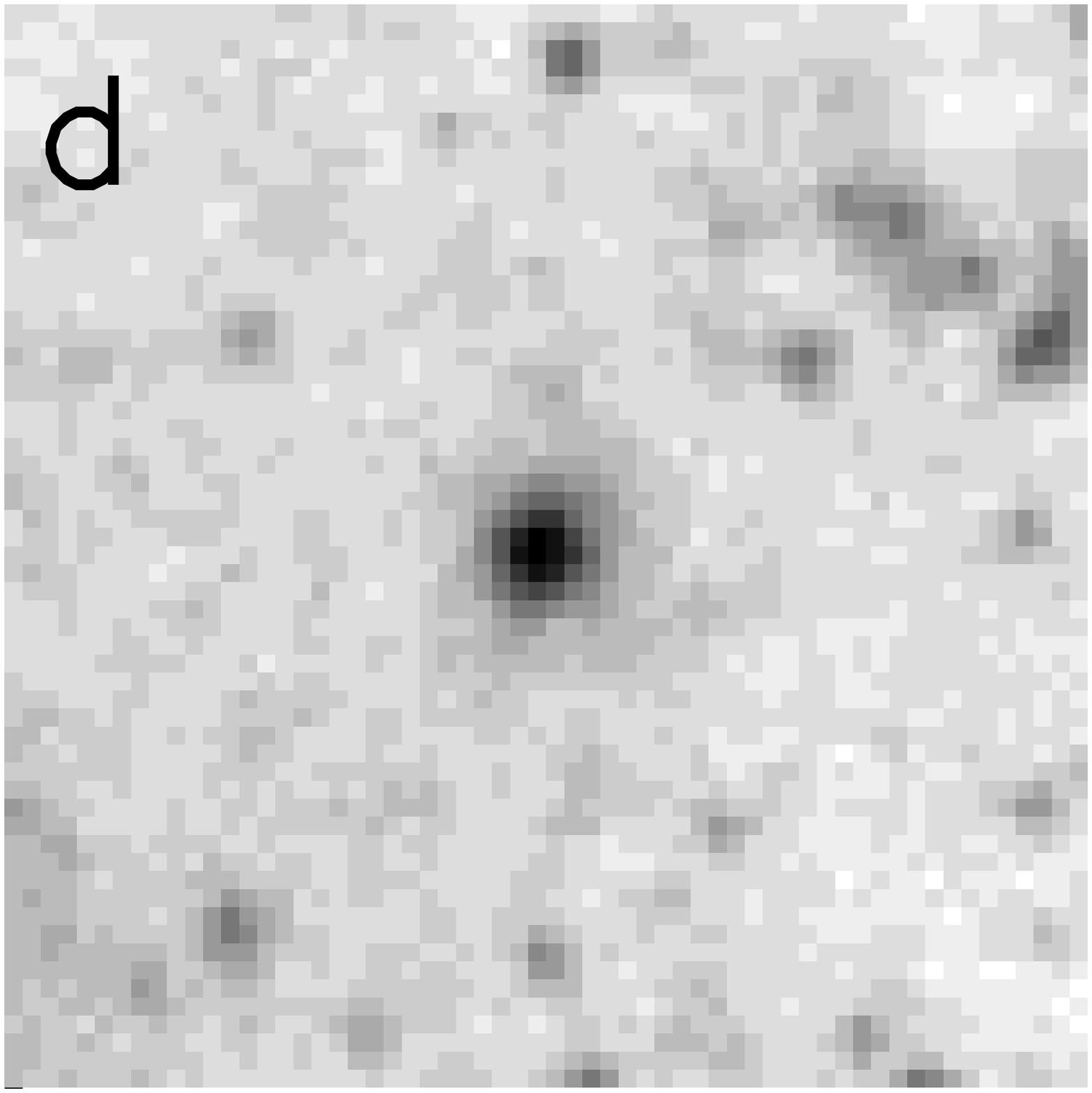}
\includegraphics[width=27mm]{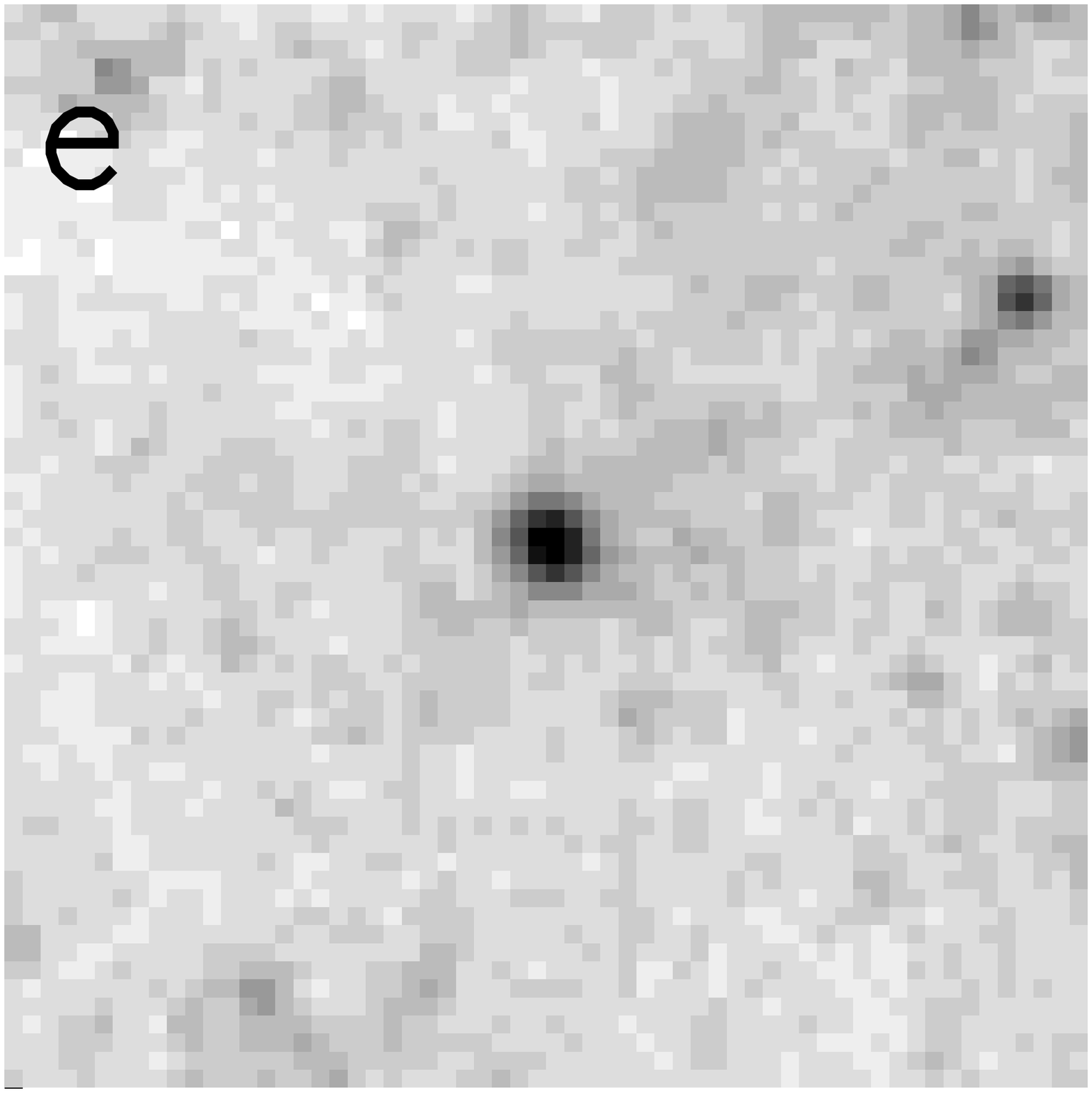}
\includegraphics[width=27mm]{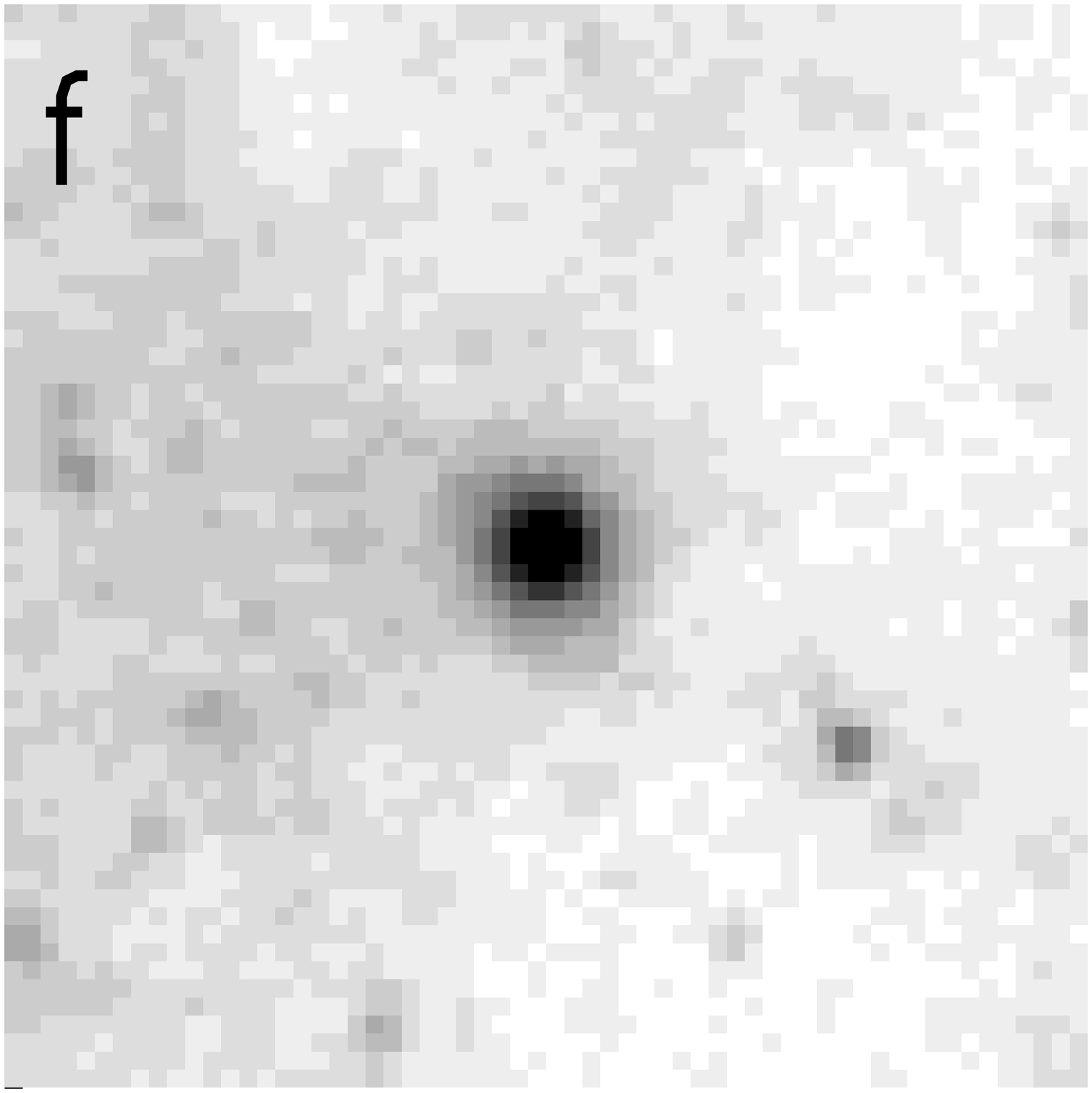}\\
\includegraphics[width=27mm]{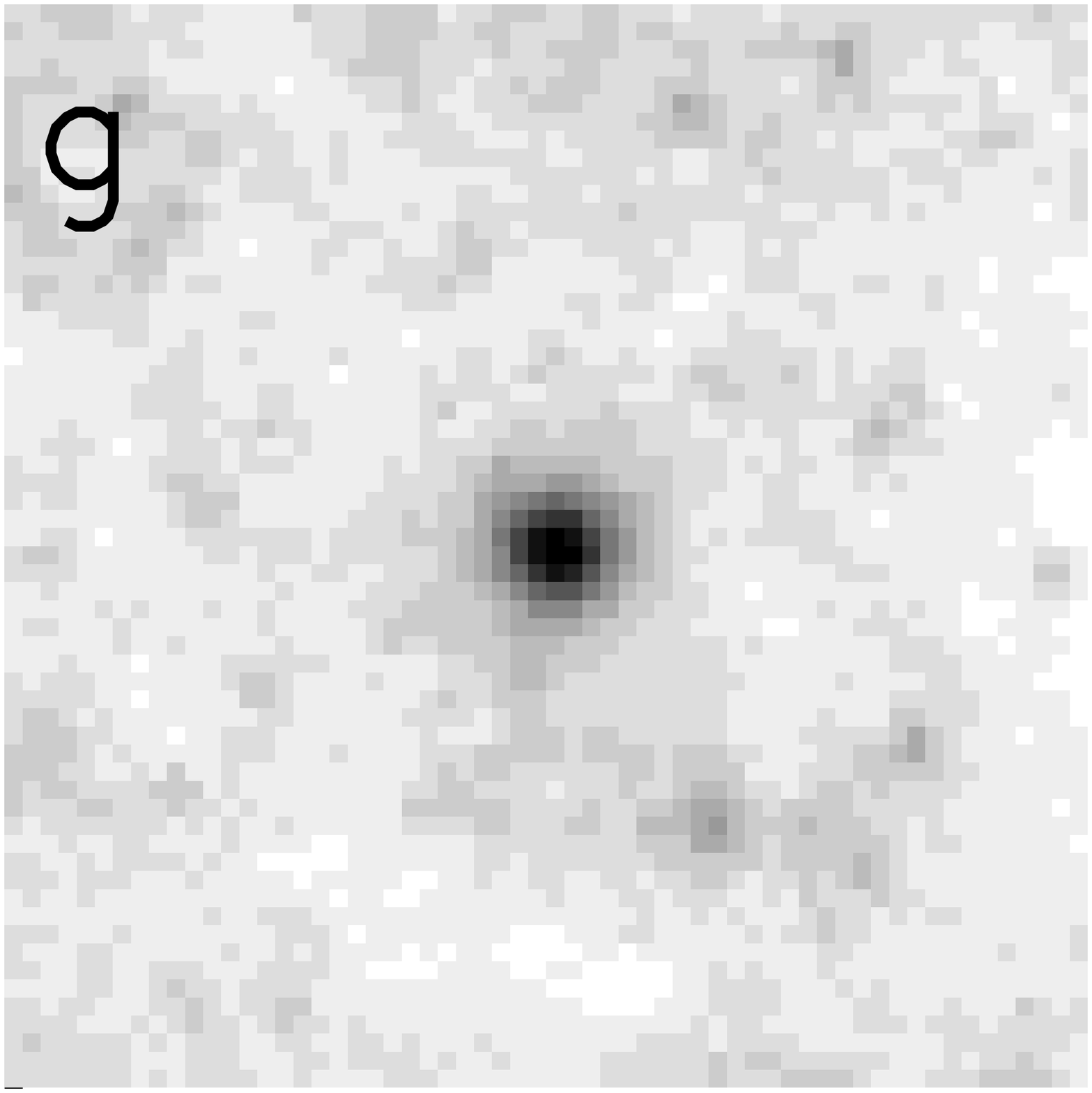}
\includegraphics[width=27mm]{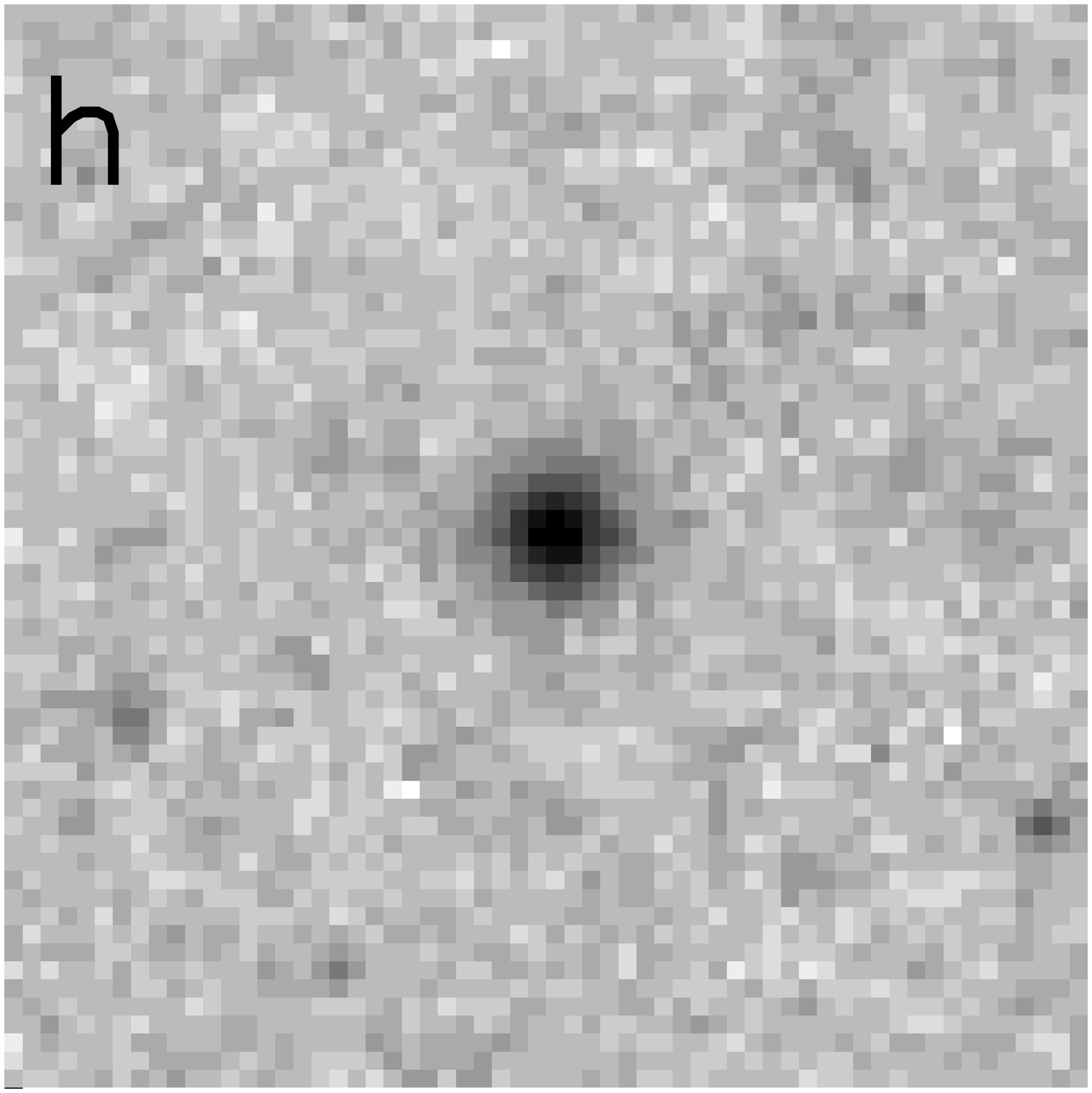}
\includegraphics[width=27mm]{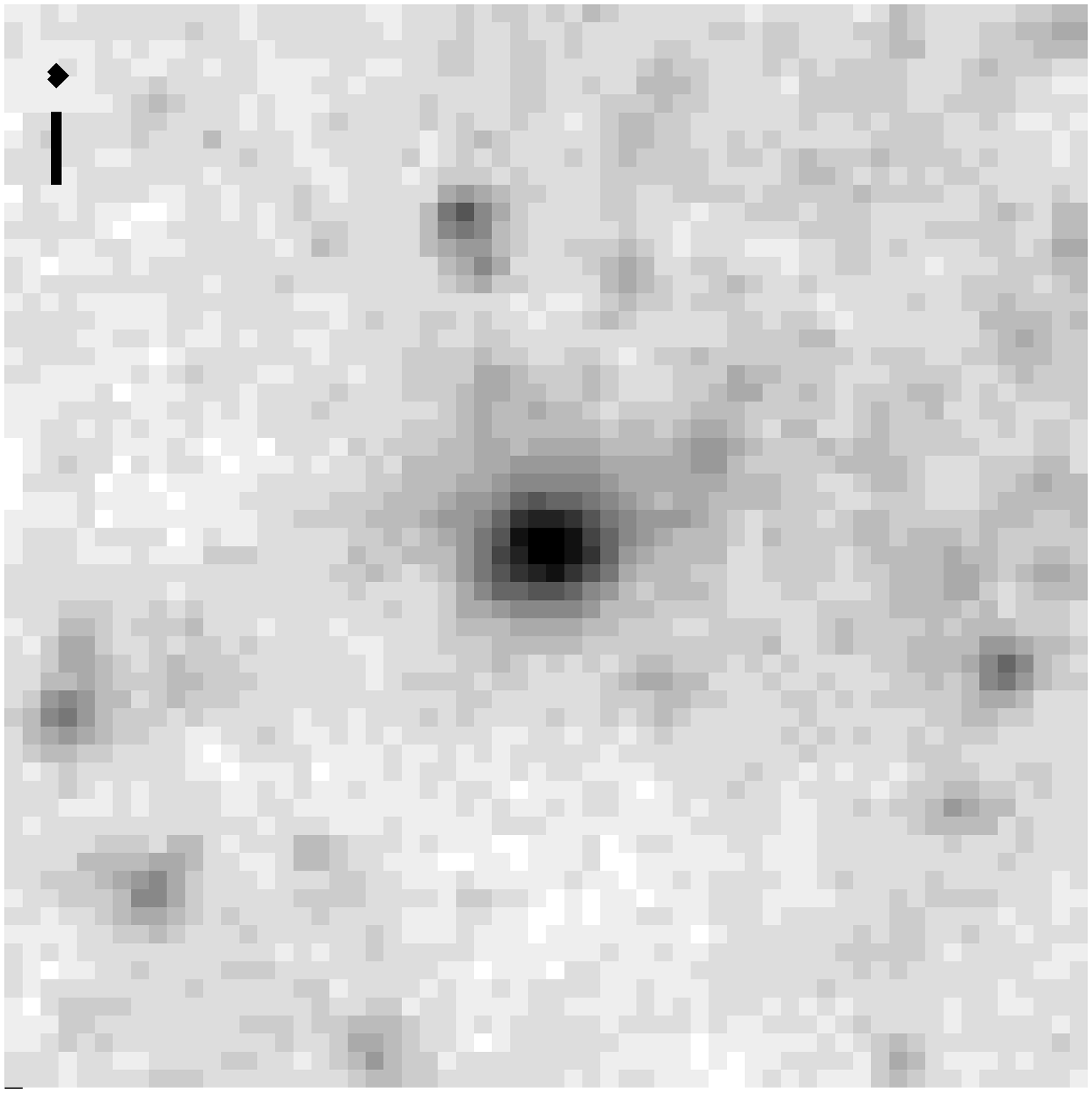}
\caption{A selection of isolated stellar cluster candidates in \bband\
that passed the visual inspection.  The size of each image is
60$\times$60 pixels or 3\arcsec$\times$3\arcsec.}
\label{fig:nice}
\end{minipage}
\end{figure}

After these criteria there were still many contaminants in our
sample, e.g.~patches of high background regions in between dark dust
lanes that were detected as a source, background galaxies, blended
sources and crowded regions.  The next criteria were used to remove
contaminants and to select only cluster candidates with accurate
radii:

\begin{enumerate}

\item[3.] Following the results of our artificial cluster experiments, we
applied the following magnitude cutoffs for our ``radius sample'':
$\bband <22.6$, $\vband <22.3$ and $\iband <22.7$.  The first two
limits were chosen according to
Fig.~\ref{fig:completeness_sizedifferences}.  With these limits for
\bband\ and \vband, more than 68\% of the artificial clusters with
input radii up to 9~pc that were retrieved, had a measured radius
better than 20\%.  In Fig.~\ref{fig:limits} we see that these limits
are brighter than the 90\% completeness limits in the high background
region for sources up to $\sim$8~pc.  Since we do not use the radii
measurements in \iband, in this passband the 90\% completeness limit
for a 3~pc source was applied.  For the selection of the larger
``resolved sample'' we used the 90\% completeness limits in all three
passbands: $\bband <24.2$, $\vband <23.8$ and $\iband <22.7$.

\item[4.] We only selected sources that had an absolute difference in
\reff\ between \bband\ and \vband\ of less than 2~pc.  Tests revealed
that mostly all sources in low, homogeneous background regions already
fulfilled this criterion and a check by eye showed that sources which
did not pass this criterion were practically all contaminants due to a
highly varying local background.
We did not apply this size difference criterion for our resolved
sample, since sources in this sample do not necessarily have accurate
radii measured in both \bband\ and \vband.

\item[5.] We used the distance of every source to its nearest neighbouring
source as a criterion to filter out unreliable fits due to blending or
crowding.  A source was rejected if it had a neighbour within 5 pixels
(the fitting radius of \ishape).  Of course this method only works
when both neighbouring sources are in the original source list.

\item[6.] To select out the remaining blended sources and crowded regions
that were initially detected as a single source, we inspected every
remaining source in our sample by eye.  We realize that this
introduces some amount of subjectivity into our sample, but we are
dealing with a face-on spiral galaxy with a high degree of
irregularities in the background light due to the spiral structure and
many crowded star forming regions.  Therefore, visual inspection was
unavoidable for our purpose of selecting a cluster sample with
accurate radii.

We created small images in all three filters for all the sources that
fulfilled the above mentioned criteria, and by visual inspection we
looked for:
\begin{itemize}
\item the presence of a second, separate peak within a distance of $\sim$5~pixels
\item a very irregular shape
\item very small fitted aspect ratio for sources that appear fairly circular
\item crowded regions
\item very steep gradients in the background light that likely influenced the fitted radius
\item background galaxies
\end{itemize}
When, based on one of these points, the source was not a clear cluster
candidate with an accurately determined radius, we rejected the source
from our sample.  In Fig.~\ref{fig:crap} we show a number of sources
that were rejected together with the main reason.  For comparison we
show a number of accepted cluster candidates in Fig.~\ref{fig:nice}.
Visual inspection removed another 24, 21 and 22\% of the sources from
our sample in the low, intermediate and high background region,
respectively.

\end{enumerate}

In Table~\ref{tab:Cluster sample} we summarize the results of our
sample selection for the radius sample.  Our final \emph{sample of
stellar cluster candidates with accurate radii} consists of 1284
sources, of which 57\% are located in the high background region and
25\% in the low background region.  From here on we will refer to this
sample of stellar cluster candidates simply as ``clusters''.  The
larger sample of \emph{resolved sources}, which does not satisfy the
radius difference criterion (criterion 4) and which magnitude cutoffs
follow the 90\% completeness limits (criterium 3), consists of 7698,
6846 and 5024 sources in \bband, \vband\ and \iband, respectively.
This sample will be studied in a different paper \citep[in prep.]{haas07}.

\begin{table}[!t]
\caption[]{The selection of the sample of stellar cluster candidates
with accurate radii. The numbers indicate the amount of sources that
satisfy the mentioned criterion and all the criteria mentioned above
it, split in the low, intermediate and high background region.}
\label{tab:Cluster sample}
\begin{center}
\begin{tabular}{lccc}
\hline\hline
Criterion & & Nr. remaining & \\
 & low & inter & high \\
\hline 
All sources & 35\,980 & 15\,809 & 23\,647  \\
$\reff> 0.5$~pc & 11064 & 4799 &  7238\\
$\chi^{2}/\chi_{0}^{2} < 1$ & 10715 & 4661 & 7028 \\
$\bband < 22.5$ mag & & & \\
$\vband < 22.3$ mag & 472 & 346 & 1301 \\
$\iband < 22.7$ mag & & & \\
$|R_{\bband}-R_{\vband}| < 2$~pc & 457 & 327 & 1068 \\
No neighbour within 5 pixels & 418 & 302 & 939 \\
After visual inspection & 317 & 239 & 728 \\
\hline
Total sample: 1284 & & &\\
\hline
\end{tabular}
\end{center}
\end{table}

\subsection{Old Globular Clusters and Faint Fuzzies}

To see if there are any possible old ($\gtrsim 10$~Gyr) GC candidates
in our cluster sample, we applied the colour criteria $\bv>0.5$ and
$\vi>0.8$ to our cluster sample, typical for old MW GCs.  There are
only 70 clusters satisfying these criteria, showing that the majority
of our cluster sample consists of young clusters, but a small fraction
of $\sim$5\% is probably part of an old GC population or highly
reddened. A more detailed identification of GC
candidates in M51 will be carried out in future studies.

\begin{figure}[t]
\centering
 \includegraphics[width=85mm]{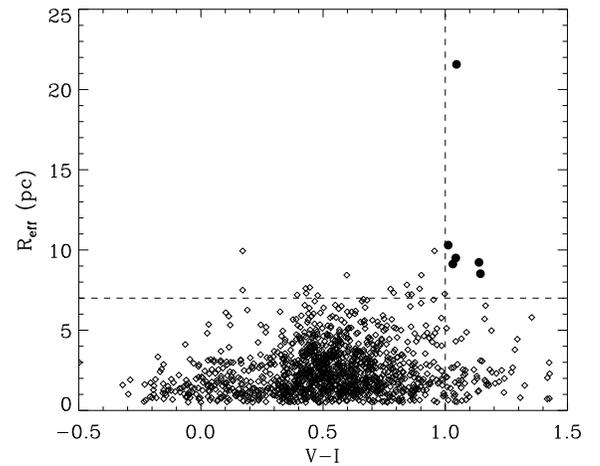}
\caption{\reff\ versus \vi\ colour for the total sample of 1284
  clusters. The upper-right corner of the dashed lines indicate
  the region of the sub-group of faint fuzzy
  candidates (filled circles). These cuts are the same as used by
  \citet{hwang06} to select faint fuzzy clusters in NGC~5195.}
\label{fig:FF}
\end{figure}

We note that our dataset covers the same field used by \citet[][from
here on referred to as ``HL06'']{hwang06}, who detect 49 ``faint
fuzzy'' star clusters around the companion of M51, NGC~5195. Faint
fuzzy clusters were introduced by \citet{larsen00a}, because they
formed a sub-group in a radius-colour diagram of star clusters in
NGC~1023. In Fig.~\ref{fig:FF} we show \reff\ versus \vi\ for the
clusters in our sample. Six clusters in our sample satisfy
the criteria of a faint fuzzy, namely $0.6<\bv<1.1$, $1.0<\vi<1.5$ and
$\reff>7$~pc. These faint fuzzy candidates are indicated in
Fig.~\ref{fig:FF}. The figure shows that the faint fuzzy candidates
indeed form a separate group in a radius-colour diagram and are not
simply the largest clusters in the tail of a continuous radius
distribution.

The six faint fuzzy candidates seem randomly located in (or projected
onto) the disc of M51. None of the 49 faint fuzzy candidates of HL06
are therefore recovered in our cluster sample.  This is because all
these 49 candidates are fainter than the magnitude limits we apply.
This is expected, since we apply rather conservative magnitude limits
in order to achieve accurate radius measurements, also for large
clusters in high background regions (\S~\ref{subsec:Selection
criteria}).  If we would drop our conservative magnitude limits from
the selection criteria, we would have 37 of the 49 faint fuzzy
candidates from HL06 in our sample. The 12 remaining candidates are
removed from our sample based on ``inaccurate radii'' criteria (large
radius difference between \bband\ and \vband\ or a $\chi^{2}$ larger
than $\chi_{0}^{2}$). The six faint fuzzy candidates in our
sample are not in the sample of HL06, because these authors focused
on the region around NGC~5195 and were therefore not covering the disc
of M51.

\subsection{The largest cluster: 212995}
  \label{subsec:The largest cluster: 212995}

\begin{figure}[t]
\centering
 \includegraphics[width=85mm]{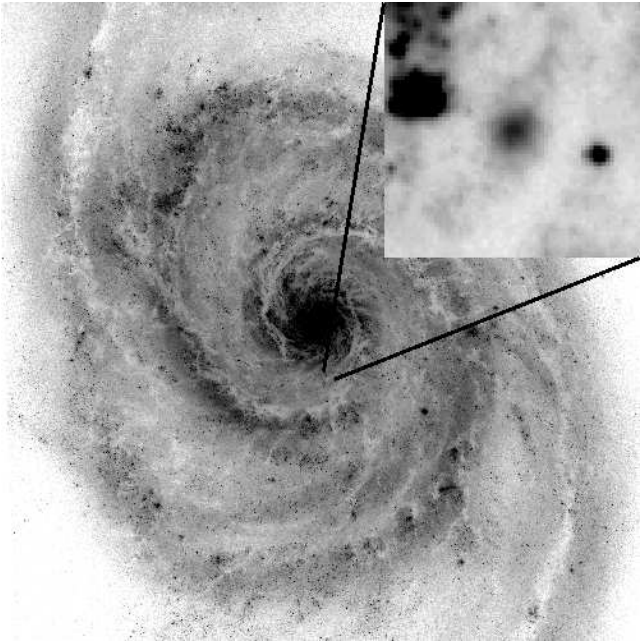}
\caption{Cluster ``212995'', the largest cluster in our sample with a
$\reff = 21.6$~pc (centre of the inset), and its location in M51
(north is up, east is to the right). The projected galactocentric
distance is 1.02~kpc. The scale of the inset image is
$\sim60\times60$~pixels or $\sim122\times122$~pc. The scale of the big
image is $\sim16\times16$~kpc. }
\label{fig:212995}
\end{figure}

One cluster candidate in our radius sample, with our ID number
``212995'', clearly stands out from the other cluster candidates
in radius.  The cluster candidate, positioned at
RA=13$^{\mathrm{h}}$29$^{\mathrm{m}}$51\fs94,
Dec=+47\degr11\arcmin19\farcs63 and shown in Fig.~\ref{fig:212995},
has an (ellipticity corrected) $\reff = 21.6$~pc in \bband.  The
projected galactocentric distance of this cluster candidate is
1.02~kpc.  Its brightness in \bband, \vband\ and \iband\ is 22.27,
21.52 and 20.48 mag, respectively, with corresponding \bv\ and \vi\ of
0.75 and 1.05, respectively.  These colours make this source both an
old GC and faint fuzzy candidate.  Assuming the source is a cluster,
we can make an estimate of its age and mass by using GALEV SSP models.
However, since we only have photometry in three filters, this estimate
suffers from an age-extinction-metallicity degeneracy, introducing
rather large uncertainties.  Assuming a low extinction ($E(B-V)< 0.1$)
and a metallicity in the range of 0.2--1.0~$\mathrm{Z_{\sun}}$, the
best estimate for the age is $2.7^{+2.8}_{-1.8}$~Gyr.  The
corresponding best estimate for the mass is
$2.5^{+3.3}_{-1.1}$$\times10^{5}~\mathrm{M_{\sun}}$, which is a lower
limit due to the underestimation of the aperture correction for the
photometry for such a large source.  Assuming the metallicity is
$0.02~\mathrm{Z_{\sun}}$, the best estimates for the age and mass are
$13.2^{+2.8}_{-9.7}$~Gyr and
$8.0^{+3.5}_{-4.6}$$\times10^{5}~\mathrm{M_{\sun}}$, respectively.
However, the possibility of the source being a highly reddened young
cluster is not fully excluded.  There is also a possibility that this
cluster candidate is actually a background galaxy, but this
possibility is considered to be unlikely, since the
cluster is located (in projection) very close to the centre of M51,
where the extinction of the background source and the surface brightness
of the foreground (M51) is high.

\section{A comparison between \acs\ and \wfpc}
  \label{sec:A comparison between ACS and WFPC2}

As a test of the robustness of our methods, we compared the photometry
and the radii of our clusters to the data of B05.  B05 derived ages,
masses, extinctions and effective radius estimates of stellar clusters
covering the central $\sim7.3\times8.1$~kpc of M51 using \hst\
\emph{NICMOS} and \wfpc\ data.  We calculated the transformation
between a mosaic of the \vband\ \wfpc\ images and the \acs\ mosaic
image with the GEOMAP task in IRAF, by identifying 10 sources by eye
that were clearly visible in both data sets.  We then transformed the
coordinates of all the clusters in their sample to our \acs\
mosaic frame using the GEOXYTRAN task in IRAF, and we matched sources
which were within 1 pixels from a cluster in our sample, which had
photometry in \bband\ ($\approx B$), \vband\ ($\approx V$) and \iband\
($\approx I$) in both data sets and which had a measured radius on the
\wfpc\ data $0 <\reff < 12$~pc.  This resulted in 271 matched
clusters, of which a few will be mismatched due to the uncertainties
in the transformation and most importantly, geometric distortion of
the \wfpc\ images.

\begin{figure}[t]
\centering
 \includegraphics[width=85mm]{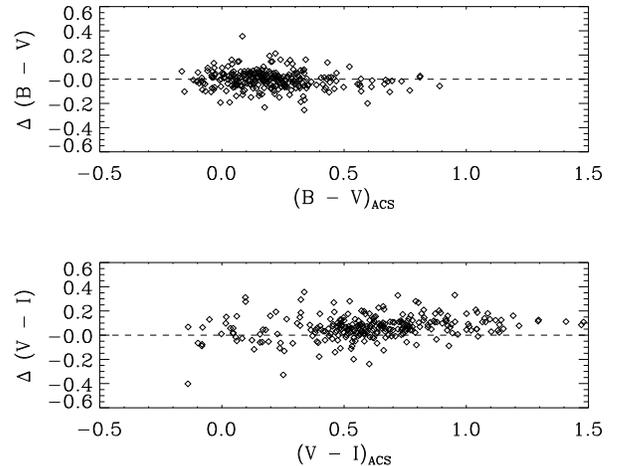}
\caption{The differences between the colours of clusters matched
between our \acs\ dataset and the \wfpc\ dataset of Bastian et al.\
(2005), versus the colour in the \acs\ dataset.The dashed lines
  show a difference of 0 as a reference for the eye.}
\label{fig:colour_acs_vs_wfpc2}
\end{figure}

For these 271 clusters we compared the \bv\ and \vi\ colours of B05
with our results, after we removed our filter dependent infinity
corrections and Galactic foreground extinction corrections, since
these were constant for the photometry of B05.  In
Fig.~\ref{fig:colour_acs_vs_wfpc2} we plot the difference between the
\acs\ and \wfpc\ colours versus the \acs\ colour.  For the mean
differences we find:
\begin{eqnarray*}
\Delta (B\!-\!V) = (B\!-\!V)_{ACS} - (B\!-\!V)_{WFPC2} & = & -0.005 \pm 0.005\\   
\Delta (V\!-\!I) = (V\!-\!I)_{ACS} - (V\!-\!I)_{WFPC2} & = & 0.064 \pm 0.006,     
\end{eqnarray*}
where the errors are the standard errors of the means, not to be
confused with the standard deviations, which are $0.076$ and $0.098$,
respectively.  This shows that our colours are in good agreement with
B05 and that we can adopt the masses derived by B05 for these 271
clusters to study the mass-radius relation with the higher resolution
of the \acs\ data in \S~\ref{subsec:Mass-radius relation}.

\begin{figure}[t]
\centering
 \includegraphics[width=85mm]{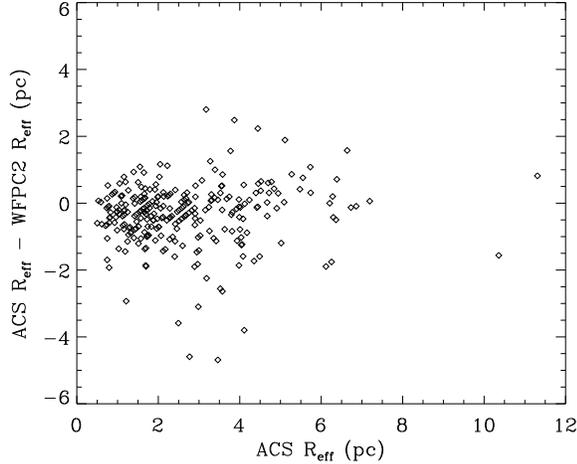}
\caption{The differences between the effective radii of the clusters
matched between our \acs\ dataset and the \wfpc\ dataset of Bastian et
al.\ (2005), versus the effective radius in the \acs\ dataset.}
\label{fig:reff_acs_vs_wfpc2}
\end{figure}

We also compared the effective radii of the 271 matched clusters on
the \vband\ image.  In Fig.~\ref{fig:reff_acs_vs_wfpc2} we show the
difference between the \acs\ and \wfpc\ radius versus the \acs\
radius.  No clear trend is visible, except that the average \acs\
radius of the clusters is slightly smaller than the \wfpc\ radius.
The mean difference between the \acs\ and \wfpc\ radius is
\begin{eqnarray*}
\reff_{,ACS} - \reff_{,WFPC2} = -0.34\pm0.06~\mathrm{pc},
\end{eqnarray*}
where the error is the standard error of the mean.  The standard
deviation around the mean is $0.94$~pc.  We note that some of the
differences between \acs\ and \wfpc\ radii are expected to be caused
by contaminants in the matching procedure, as well as resolution
effects (blends in the \wfpc\ data) and a different treatment of
ellipticity for both data sets.  For the \wfpc\ data circular cluster
profiles were assumed, while for the \acs\ data we used elliptical
profiles with a transformation to a single \reff.
Overall, the mean difference between the \acs\ and \wfpc\ radii is
within the expected accuracy of the radius measurements
($\sim0.5$~pc), and Fig.~\ref{fig:reff_acs_vs_wfpc2} suggests that
there are no strong radius dependent biases in our methods.

\section{The radius distribution}
  \label{sec:Radius distribution}

Now that we have selected a sample of clusters with accurate radii, we
will study the distribution of their radii and possible dependencies
between radius, background region and colour in this section.
Possible correlations between radius and luminosity, mass and
galactocentric distance will be the subject of the next section.

We show the effective radius distribution\footnote{Strictly speaking,
the term \emph{distribution} refers to linear intervals, i.e.~$N(R)\ud
R$, and the term \emph{function} refers to logarithmic intervals,
i.e.~$N(R)\ud \mathrm{log}R$. In this work, however, we will not make
this distinction. We will only use the term \emph{distribution} and we
will specify the type of interval used when necessary.} of our sample
of 1284 clusters with linear bins in Fig.~\ref{fig:Radius distribution
linear} and with logarithmic bins in Fig.~\ref{fig:Radius distribution
loglog}. In both figures we plot the radius distributions for cluster
in the low and high background region separately.  We see that the
radius distribution peaks around 1--2~pc and then drops to a maximum
radius of $\sim$10~pc. In the remainder of this section we will first
focus on the slope of the radius distribution at $\reff>3$~pc and then
focus on the location of the peak.

\begin{figure}[t]
\centering
\includegraphics[width=85mm]{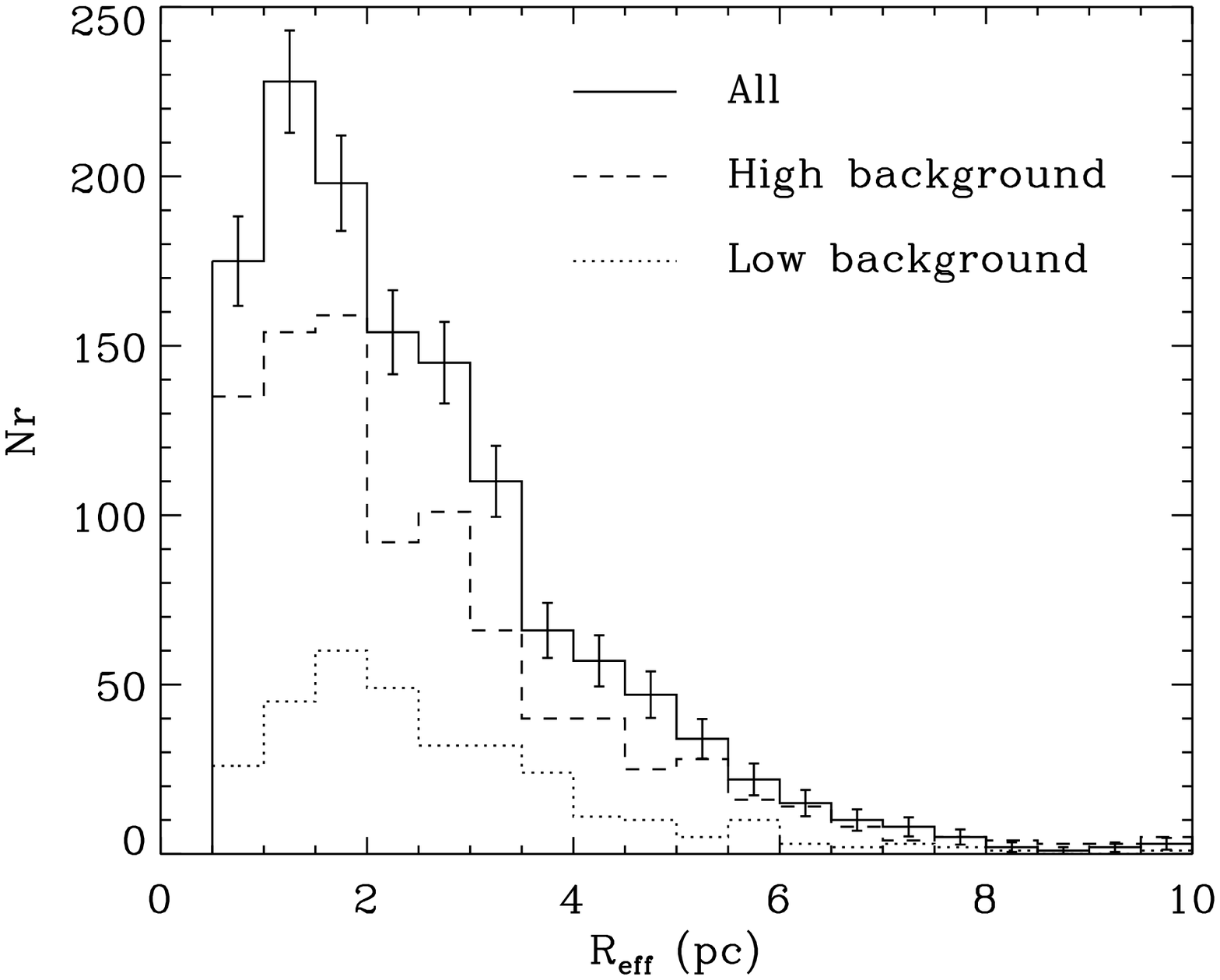}
\caption{The effective radius distribution of the total sample of 1284
clusters, measured on the \bband\ image, using linear radius bins
(solid line). Also shown are the radius distribution of only the
sources in the low background region (dotted line) and the sources in
the high background region (dashed line). For better readability, only
the (Poissonian) error bars of the total sample are shown.}
\label{fig:Radius distribution linear}
\includegraphics[width=85mm]{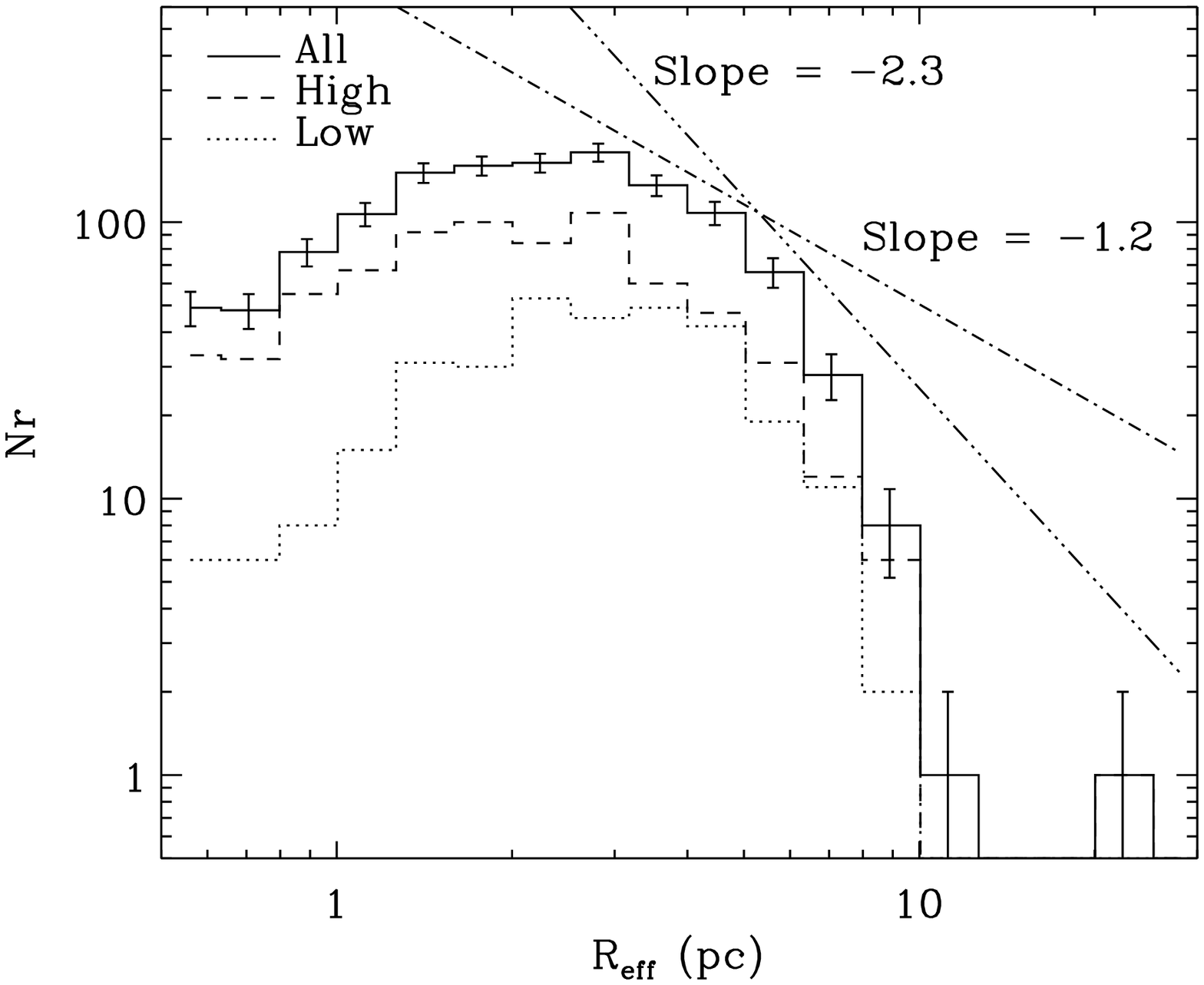}
\caption{The effective radius distribution of the total sample of 1284
clusters, measured on the \bband\ image, using logarithmic radius bins
(solid line). Also shown are the radius distribution of only the
sources in the low background region (dotted line) and the sources in
the high background region (dashed line). For better readability, only
the (Poissonian) error bars of the total sample are shown. For
comparison we also show a power law with a slope of $-2.3$, typical
for a fractal radius distribution of star-forming gas clouds, and a
slope of $-1.2$ as determined by B05 for the radius distribution of
407 star clusters in M51 with \wfpc\ data.}
\label{fig:Radius distribution loglog}
\end{figure}

\subsection{The slope of the radius distribution}
  \label{subsec:The slope of the radius distribution}

It has been observed that the \emph{mass} distributions of both
star-forming clouds \citep{sanders85, solomon87, harris94, brand95,
elmegreen96a, fukui01} and star clusters \citep{zhang99, ashman01,
bik03, degrijs03a, hunter03} can be approximated by power laws of the
form $N(M)\ud M \propto M^{-\alpha}\ud M$, with the index $\alpha$ in
the range of 1.5--2.0.  Star-forming clouds also show a power-law
\emph{radius} distribution of the form $N(R) \ud \log R \propto
R^{-2.3}\ud \log R$ \citep{harris94, elmegreen96a}. For the clouds,
the power-law mass and radius distributions are consistent with the
clouds having a fractal structure with a fractal dimension of 2.3
\citep{mandelbrot83, elmegreen01b}. Since the mass distribution of
clusters is similar to the mass distribution of clouds, one might
naively expect the radius distributions also to be similar.  We
plotted the radius distribution of star clusters in M51 in
Fig.~\ref{fig:Radius distribution loglog}, using logarithmic bins. In
this figure a power law would be a straight line. We plotted two lines
with a slope of $-2.3$ and $-1.2$.  The first slope is consistent with
the power-law distribution of the form $N(R) \ud \log R \propto
R^{-2.3}\ud \log R$ or $N(R) \ud R \propto R^{-3.3}\ud R$, observed
for star-forming gas clouds on every length scale (down to the
smallest scales of $\sim0.1$~pc).  We see in Fig.~\ref{fig:Radius
distribution loglog} that the radius distribution of the clusters in
M51 between $\sim$3 and $\sim$10~pc can \emph{not} be approximated by
the same power law as the one for the star-forming gas
clouds.\footnote{Between 0.5 and $\sim$10~pc a log-normal
    distribution provides a reasonable fit to the data (not shown here).}

The slope of $-1.2$, indicated in Fig.~\ref{fig:Radius distribution
loglog}, is consistent with the power-law distribution of $N(R) \ud R
\propto R^{-2.2}\ud R$, found by B05 for 407 clusters between 2--15~pc
in M51 using \wfpc\ data.  Although the slope of our observed radius
distribution in the range $\sim3$--6~pc is similar to the slope
observed by B05, our observed radius distribution is considerably
steeper at larger radii. However, we note that we used a larger
sample of clusters, measured at twice the resolution and which was
checked by visual inspection for contaminants and blends. The cluster
sample of B05 is therefore expected to have a larger fraction of
contaminants and blends than the current sample. We note however,
that the current sample is still biased against clusters in crowded
regions, but for the remainder of this study we assume that the
current sample is representative of the entire star cluster
population of M51.

Fig.~\ref{fig:Radius distribution loglog} shows that the radius
distribution of star clusters in M51 is not consistent with a fractal
structure. This suggests that after the formation of the clusters from
a fractal gas, their radii have changed in a non-uniform
way. Unfortunately, at the small radius end of the distribution a
possible bias against small clusters can not be completely ruled out,
since in a fractal gas the \emph{smallest} clusters are expected to form in
groups \emph{closest} together. These small clusters could have been rejected
from our sample by the close neighbour criterion
(\S~\ref{sec:Selection of the sample}). Without this criterion,
however, our sample would also be biased against small clusters due to
blends. This bias is not expected at the large radius end of the
distribution, where the radius distribution of star clusters in M51 is
clearly not consistent with a fractal distribution.

The picture in which the radii of the clusters change
shortly after their formation is consistent with various suggested
explanations for the lack of the mass-radius relation of clusters (see
\S~\ref{subsec:Mass-radius relation}).  One possible explanation is
that interactions between young star clusters and gas clouds lead to
dynamical heating and therefore expansion of the clusters
\citep[e.g.][]{gieles06c}. This expansion will be strongest for the
largest and least concentrated clusters because of their lower density and it will therefore
make an initial power-law distribution more shallow.  Although
cluster-cloud interactions are probably part of the explanation for
the lacking mass-radius relation, on its own this scenario fails to
explain the expansion of the smallest clusters, which is necessary to
weaken the mass-radius relation.

Another suggested explanation for the weak or missing mass-radius
relation of clusters is a star formation efficiency (SFE) which scales
with the mass of the clouds \citep{ashman01} combined with the early
residual gas loss from clusters \citep{hills80, geyer01, goodwin06}. In this
scenario the forming clusters initially follow the same mass-radius
relation as the clouds.  However, the removal of binding energy will
lead to the expansion of preferably small clusters, since they form
from low-mass clouds which lose relatively more residual gas.  On its
own, however, this explanation will lead to a steeper radius
distribution of clusters compared to clouds, i.e. with a slope
$<-2.3$, contrary to what we observe (Fig.~\ref{fig:Radius distribution loglog}).

Although Fig.~\ref{fig:Radius distribution loglog} suggests that the
radii of the clusters have changed shortly after their formation, our
radius distribution is consistent with neither scenario. Perhaps a
scenario including a combination of stochastic cluster-cloud
interactions, expansion due to residual gas expulsion and a
mass-dependent SFE can change the radius distribution in a way
that is more consistent with the observed radius
distribution. However, a fundamental problem of the missing
mass-radius relation of clusters is that there are also high-mass
clusters with \emph{small} radii. The existence of these clusters can
not be explained by the scenarios mentioned so far, which all rely on
the expansion of clusters. Therefore, we need better scenarios and more
insight in cluster formation theory to account for the differences in the radius
distributions between clouds and clusters.

\subsection{The peak of the radius distribution}
  \label{subsec:The peak of the radius distribution}

In Fig.~\ref{fig:Radius distribution linear} we see that there is a peak
in the radius distribution around $\sim$1.5~pc. If we assume that
the star clusters in M51 formed from a fractal gas, this is consistent
with the suggestion of the expansion of preferably the smallest
clusters, i.e.\ cluster with $\reff < 1.5$~pc which expanded to radii $>$1.5~pc. 

Fig.~\ref{fig:Radius distribution linear} and \ref{fig:Radius
distribution loglog} also show that the radius distribution of
clusters in the low background region (the dotted lines) shows a more
pronounced turnover, i.e. there are relatively fewer small clusters in
the low background region compared to the high background region.
This suggests that generally the smallest clusters are mainly found in
the high background regions (e.g.~inside the spiral arms). The
medians also show this trend: while the median \reff\ of our complete
cluster sample is 2.1~pc, it is 1.9~pc for the high background and
2.7~pc for the low background region.

We stress that it is \emph{not} very likely that this trend is biased
due to selection effects, since we applied magnitude limits that are
brighter than the 90\% completeness limits in the high background
region (\S~\ref{sec:Artificial cluster experiments}), and visual
inspection removed most background galaxies in the low background
region and possible blends in the high background region. Also, the
more compact clusters are easier to detect, so it is not likely that
there is a selection effect against small clusters in the low
background regions.

\begin{figure}
\centering
 \includegraphics[width=85mm]{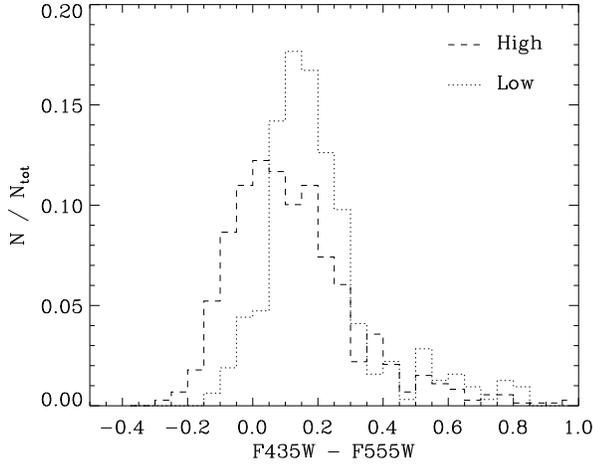}
\caption{The normalized \bv\ colour distribution for clusters in the low (dotted) and high (dashed) background region. The high background region has a higher fraction of blue clusters than the low background region.}
\label{fig:Background_vs_color}
\end{figure}

Fig.~\ref{fig:Background_vs_color} shows the \bv\ colour distribution
of clusters in the low and high background region.  The high
background region has a higher fraction of blue clusters ($\bv\lesssim
0.05$) than the low background region. This fraction is expected to
be stronger when extinction is taken into account, since clusters in
the high background region are likely more strongly reddened than
clusters in the low background region.

\begin{figure}
\centering
 \includegraphics[width=85mm]{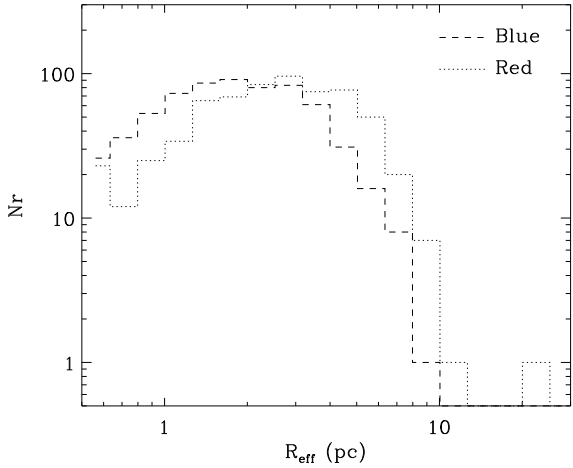}
\caption{The effective radius distribution split in a sample with
$\bv < 0.1$ (blue, dashed line, 645 clusters) and $\bv > 0.1$ (red,
dotted line, 639 clusters). The radius distribution of the red sample
is shifted towards larger radii.}
\label{fig:Radius distribution color}
\end{figure}

\begin{figure}
\centering
 \includegraphics[width=85mm]{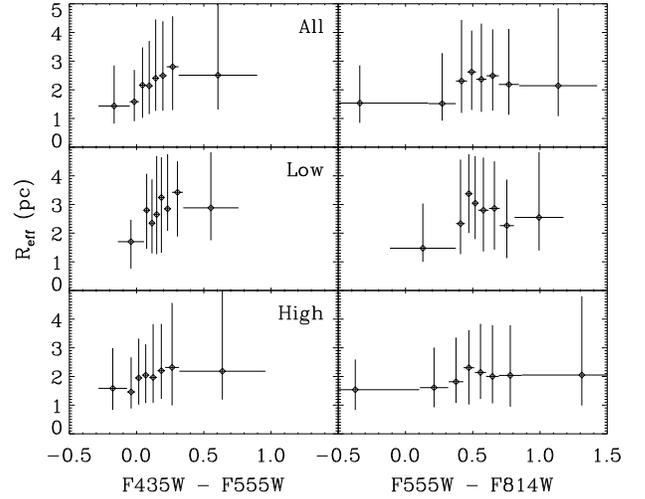}
\caption{The median \reff\ versus colour for bins containing an equal
number of clusters for the complete cluster sample (top) and the low
(middle) and high (bottom) background region.  The vertical error bars
indicate the range in which 68\% of the clusters is contained.}
\label{fig:Radius_vs_color}
\end{figure}

Fig.~\ref{fig:Radius distribution loglog} and~\ref{fig:Background_vs_color}
suggest that there is a relation between \reff\ and colour.  In
Fig.~\ref{fig:Radius distribution color} we show the radius
distribution for 2 equal-sized samples with $\bv < 0.1$ (``blue'') and
$\bv > 0.1$ (``red'').  We indeed see a shift in the radius
distribution towards larger radii for our red subsample.  The median
value follows this shift: for the blue sample the median \reff\ is
1.8~pc, while for the red sample it is 2.5~pc.  In
Fig.~\ref{fig:Radius_vs_color} we show the median \reff\ versus \bv\
and \vi\ colour.  Because all bins contain an equal number of
clusters, size-of-sample effects are excluded.  Both for low and high
background regions and \bv\ and \vi\ colours we see a similar trend of
a median \reff\ increasing with colour, although the scatter is high
and the trend is strongest for \bv\ colours.

Therefore, the observed difference in the radius distribution between
low and high background regions can be explained by a higher fraction
of red clusters in low background regions, which are generally
slightly larger.  For young clusters, colours become redder with age.
This is consistent with a larger fraction of blue clusters in high
background regions, since these regions follow the high density spiral
arms, where most clusters are expected to form. If the observed
spread in colour is also a spread in the age of the clusters, the
slight increase in median \reff\ with colour in
Fig.~\ref{fig:Radius_vs_color} suggests a dynamical evolution of the
clusters with age. The fact that the increase in radius is
strongest for \bv\ colours supports this suggestion, because \bv\ is
more sensitive to age than \vi.

In this hypothesis, newly formed clusters in spiral arms are generally
small, reflecting the high pressure and density of their parental gas
clouds.  In the subsequent early evolution of the clusters an increase
in size is expected, likely due to dynamical heating from
cluster-cluster and cluster-cloud encounters and due to the removal of
binding energy when the clusters lose mass \citep{goodwin97,
boily03}. Clusters also expand when moving out of the spiral arm, due
to tidal forces from the spiral density wave \citep{gieles07}. This
hypothesis is consistent with the low background regions containing a
relatively larger fraction of older, more evolved clusters with
therefore slightly larger radii.\footnote{If one would actually quantify any
age-radius relation one needs to be aware of possible biases, due to a
slight mass-radius relation or size-of-sample effects. E.g.\ at older
ages, the low-mass clusters will first fade below the detection limit,
so any observed age-radius relation could then result from a possible
mass-radius relation. Also, if one would let the absolute age
intervals increase with age (i.e. logarithmic binning), one would
sample the radius distribution up to larger radii for older ages and
the average radius would seem to increase with age.}

If clusters expand, they will do this on a dynamical (crossing)
timescale of a few Myrs \citep{lada03}.  The typical timescale for a cluster
to move from the high to low background region will be about half the
time between 2 spiral arm passages, which for a 2-armed spiral galaxy
is
\begin{equation}
\tau(R_{\mathrm{G}}) = \frac{\pi R_{\mathrm{G}}}{2\left(V_{\mathrm{disc}}-\Omega_{\mathrm{p}}R_{\mathrm{G}}\right)},
\end{equation}
in which $V_{\mathrm{disc}}$ is the circular velocity in the disc and
$\Omega_{\mathrm{p}}$ is the angular pattern speed.  For M51 this
gives $\tau(1~\mathrm{kpc}) \approx 10$~Myr (using $V= 200~\kms$
\citep{garcia-burillo93} and $\Omega_{\mathrm{p}} = 37~\kms
\mathrm{kpc}^{-1}$ \citep{zimmer04}).  This is a typical lower-limit
for the timescale to move from the high to low background region. For
the more average galactocentric distance of 5~kpc,
$\tau(5~\mathrm{kpc}) \approx 5.2\cdot 10^{8}$~yr. These timescales
are longer than the expansion timescale of the clusters, and are
therefore consistent with the low background region containing a
considerable number of larger clusters than the high background
region, if the clusters started expanding shortly after their
formation in a spiral arm.

\subsection{A radius-age relation?}
  \label{subsec: A radius-age relation?}

We have used the 271 matched clusters with an age estimate from B05 to
search for a correlation between age and \reff.  These clusters show a
weak relation between radius and age of the form $\reff \propto
\mathrm{age}^{0.08\pm0.03}$, with a large scatter.  This is consistent
with the relation \citet{lee05} observe for clusters in M51 using
\wfpc\ data (best fit slope of $0.06\pm0.02$).  Fig.~\ref{fig:Radius
distribution age} shows the radius distribution for the matched
clusters, split in two equal-sized samples with log(age)$<$7.5 and
log(age)$>$7.5.  The distributions are not very different, although a
weak trend is visible since the older sample has slightly more large
clusters than the younger sample.  However, these differences are
within the statistical errors and also a K-S test did not give a
significant result ($p = 0.126$).  The median \reff\ follows a similar
trend: the median \reff\ is 1.8 and 2.2~pc for the younger and older
population, respectively.

\begin{figure}
\centering
 \includegraphics[width=85mm]{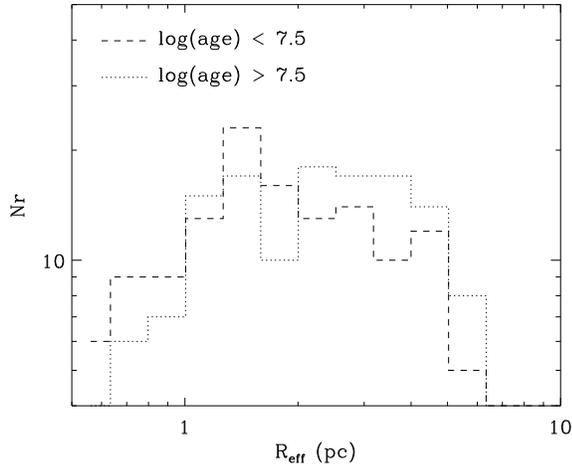}
\caption{The radius distribution of the 271 matched clusters for which we have age estimates from B05, split in a sample with log(age)$>$7.5 and log(age)$<$7.5.}
\label{fig:Radius distribution age}
\end{figure}

\subsection{A comparison with other results}
  \label{A comparison with other results}

We compared the mean and median \reff\ of our cluster sample to other
work, but we note that these comparisons can easily be biased by
differences in the lower limits of the radius and other selection
criteria.  The mean and median \reff\ of our total sample are 2.5 and
2.1~pc, respectively.  The mean and median \reff\ of the 70 GC
candidates in our sample are $3.7\pm0.4$ and 2.4~pc, respectively,
where $0.4$ is the standard error of the mean (the standard deviation
is 3.4~pc).  If we restrict ourselves to clusters in the low
background regions, the mean and median \reff\ are 2.9 and
2.7~pc, respectively. This last value is the same as the mean \reff\
\citet{jordan05} found for thousands of GCs observed in 100 early-type
galaxies of the \acs\ \emph{Virgo Cluster Survey}.
\citet{larsen04b} studied the effective radii of stellar clusters in
18 nearby spiral galaxies using \hst\ \wfpc\ images, and he found a
mean \reff\ of $3.94\pm0.12$~pc.  \citet{lee05} studied the radii of
clusters in M51 using \hst\ \wfpc\ observations covering parts of the
mosaic image used in our study, and they found a mean and median
\reff\ of 3.7 and 3.1~pc, respectively. The \acs\ camera has about
twice the resolution of the \wfpc\ camera and can therefore detect
small clusters more efficiently. This could explain why our mean and
median \reff\ are smaller than the values from \citet{lee05}. The
median \reff\ for Galactic GCs is $\sim$3.3~pc \citep{harris96}, very
similar to the value \citet{barmby06} found for their sample of blue
clusters in M101, namely 3.2~pc.  We see that the mean and median
values of \reff\ of our cluster sample are somewhat smaller than what
is generally found, although the differences become smaller when we
restrict ourselves to the clusters in the low background regions or
the GC candidates.

\section{Correlations between mass, radius and distance}
  \label{sec:Correlations between mass, radius and distance}

In this section we will combine the effective radii of the clusters
with other parameters, such as their galactocentric distance,
luminosity and for some clusters their mass.  Correlations between
these parameters for clusters in M51 have already been studied by B05,
using \wfpc\ data of the inner 5~kpc of the disc of M51.  We again
search for correlations, but now using the \acs\ data out to a
galactocentric distance of $\sim$10~kpc.

\subsection{Galactocentric distance}
  \label{subsec:Galactocentric distance}

For Galactic GCs there is a relation between the distance to the
Galactic centre, $R_{\mathrm{G}}$, and the half-light diameter
$D_{0.5}$, of the form $D_{0.5}\propto R_{\mathrm{G}}^{1/2}$
\citep{vandenbergh91}.  This same trend is observed for the core radii
of old clusters in the LMC \citep{hodge62, mateo87} and for the sizes
of old GCs in NGC 5128 \citep{hesser84}.  However, these cluster
populations are all old and mainly residing in the outer regions or
halo of their host galaxies, while we are studying a population of
mostly young clusters in a disc.

For the GCs, different explanations for the observed relation between
radius and galactocentric distance have been suggested.  One
possibility could be that the sizes of GCs reflect the densities of
the gas clouds from which they formed, i.e.\ compact GCs
preferentially formed from dense gas clouds near the centres of
galaxies, while larger GCs preferentially formed in the less dense
halo regions \citep{vandenbergh91}.  
\citet{harris94} use the
Ebert-Bonnor relations \citep{ebert55, bonnor56} to show that
\begin{equation} \label{eq:Ebert-Bonnor}
R_{\mathrm{GMC}} \propto M_{\mathrm{GMC}}^{1/2}P_{s}^{-1/4},
\end{equation}
in which $R_{\mathrm{GMC}}$, $M_{\mathrm{GMC}}$ and $P_{s}$ are the
radius, mass and surface pressure of the gas clouds, respectively.
The Van den Bergh relation then arises naturally if the clusters form
from gas clouds of which the surface pressure scales with the ISM
pressure ($P_{s}\approx 5$--$10 P_{ISM}$), which in turn scales as
\begin{equation} \label{eq:ISM}
P_{ISM} \propto R_{\mathrm{G}}^{-2}
\end{equation}
for the halo region \citep{harris94}.  From Eq.~\ref{eq:Ebert-Bonnor}
and \ref{eq:ISM} and the observation that the mean GC mass does not
change with galactocentric distance \citep{harris94}, the Van den
Bergh relation follows.  However, models like these assume that the
relations with surface pressure are also valid in the cloud cores,
where the clusters actually form, but this assumption is not
necessarily valid.

Another possible explanation for the observed relation between radius
and galactocentric distance for GCs is more evolutionary in nature.
It assumes that the GCs have reached tidal equilibrium with their host
galaxy.  The tidal radius of a cluster in an external logarithmic
potential field scales as:
\begin{equation} \label{eq:baumgardt}
r_{\mathrm{t}} \propto M_{\mathrm{cl}}^{1/3}R_{\mathrm{G}}^{2/3},
\end{equation}
where $M_{\mathrm{cl}}$ is the cluster mass \citep[][chap.\
7.3]{BT87}.  Thus, when a cluster is relaxed, in tidal equilibrium
with its host galaxy and filling its Roche lobe, its tidal radius is
expected to scale as $r_{\mathrm{t}}\propto R_{\mathrm{G}}^{2/3}$.  We
would also expect the effective radius to scale as $\reff\propto
R_{\mathrm{G}}^{2/3}$, \emph{only} if the density profile of the
cluster would be constant and if the cluster is in tidal equilibrium
with the galaxy.  For young clusters in the disc, however, the
validity of these assumptions remains to be seen.

\begin{figure}
\centering
 \includegraphics[width=85mm]{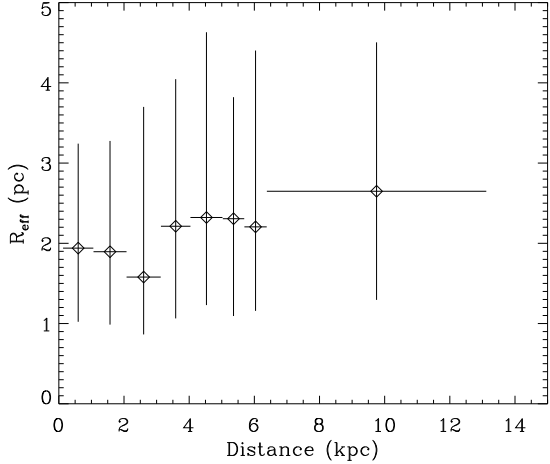}
\caption{The median \reff\ versus galactocentric distance, for bins
containing an equal number of clusters. The vertical error bars
indicate the range in which 68\% of the clusters is contained.}
\label{fig:Radius_vs_distance}
\end{figure}

In Fig.~\ref{fig:Radius_vs_distance} we show the median \reff\ versus
the galactocentric distance for the clusters in M51.  There seems to
be a slight increase in \reff\ with distance, but the scatter is large
(reflected in the large error bars).  We have tried to fit a function
of the form
\begin{equation}
\frac{\reff}{\mathrm{1~pc}} = c \cdot \left(\frac{R_{\mathrm{G}}}{\mathrm{1~kpc}}\right)^{\alpha}
\end{equation}
to the unbinned data, in which $c$ is a constant, and for the best fit
we find $\alpha = 0.12\pm0.02$.  This relation is significantly weaker
than the predicted $\alpha = 0.67$ (for GCs in tidal equilibrium) or
the observed relation for Galactic GCs ($\alpha = 0.5$).

\begin{figure}
\centering
 \includegraphics[width=85mm]{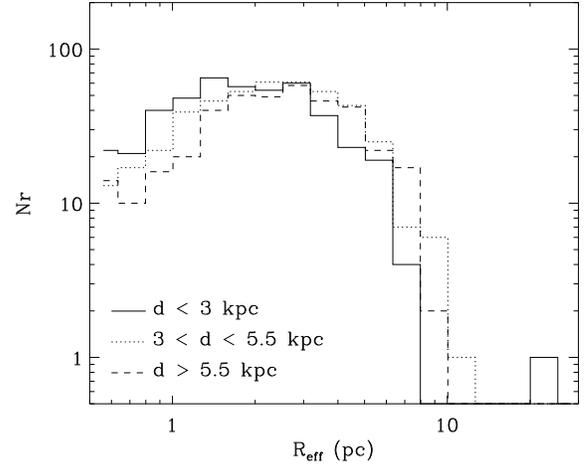}
\caption{The effective radius distributions for three different
galactocentric distance regions in M51. The solid line shows clusters
close to the centre of M51 (distance $<$~3~kpc), the dotted line shows
clusters at intermediate distances (3$<$~distance~$<$5.5~kpc) and the
dashed line shows clusters at large galactocentric distances
($>$~5.5~kpc). The changes in the distributions are consistent with a
slight increase in \reff\ with distance.}
\label{fig:Radius distribution distance}
\end{figure}

The radius distribution changes for different galactocentric distance
bins, as shown in Fig.~\ref{fig:Radius distribution distance}.  The
radius distribution close to the centre of M51 (distance $<$~3~kpc) is
shifted towards smaller radii compared to the radius distributions at
larger distances.  A K-S test confirmed that it is unlikely ($p <
0.001$) that the radius distribution for distance $<$~3~kpc and
$>$~5.5~kpc are drawn from the same parent distribution.

\begin{figure}
\centering
 \includegraphics[width=85mm]{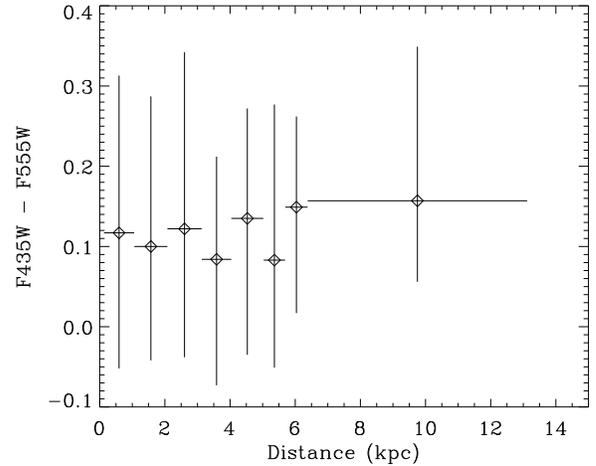}
\caption{The median \bv\ colour versus galactocentric distance of the
1284 clusters in our sample for bins containing an equal number of
clusters. The vertical error bars indicate the range in which 68\% of
the clusters is contained. No obvious trend is visible in the data.}
\label{fig:Color_vs_distance}
\end{figure}

Since we observe a relation between \reff\ and color
(Fig.\ref{fig:Radius_vs_color}), any increase in radius with
galactocentric distance could be the result of an increase in colour
with galactocentric distance.  In Fig.~\ref{fig:Color_vs_distance} we
show \bv\ versus galactocentric distance for the 1284 clusters that
were also shown in Fig.~\ref{fig:Radius_vs_distance}.  No obvious
trend is visible, which is expected since at all galactocentric radii
we encounter clusters in both arm and interarm regions.  The arm
regions are mostly high background regions and will therefore contain
mostly blue clusters, while the interarm regions are mostly low
background regions and will therefore contain mostly red clusters
(Fig~\ref{fig:Background_vs_color}).  The observed $\reff\propto
R_{\mathrm{G}}^{0.12\pm0.02}$ is therefore not likely a result of the
relation between radius and colour.

B05 did not find a relation between \reff\ and $R_{\mathrm{G}}$ for
$R_{\mathrm{G}} < 5$~kpc.  For $R_{\mathrm{G}} \lesssim 13$~kpc we
find a weak relation, but this relation is considerably weaker than
the observed relations for old GCs.  Therefore, the clusters we
observe in the disc of M51 are either forming under different
conditions than the GCs, or the observed relation for GCs emerged
during their longer dynamical evolution. We consider the first
explanation to be the most likely one, since GCs probably formed
outside a spiral disc, in regions where the surface pressure of their
parental clouds decreased with distance (Eq.~\ref{eq:ISM}). For
clusters in spiral arms it is not expected that the surface pressure
scales in a similar way with distance. Likely, the higher pressure
inside spiral arms decreases less strongly with galactocentric
distance. \citet{rix93} find that the arm/interarm density
contrast for M51 increases with galactocentric distance, consistent
with this picture. In this case
a strong radius-distance correlation for the clusters is not expected.

\subsection{Mass-radius relation}
  \label{subsec:Mass-radius relation}

One of the most peculiar properties of star clusters is the lack of a
clear relation between their mass and radius.  Star clusters are
believed to form from Giant Molecular Clouds (GMCs), for which a clear
relation between mass and radius is observed.  \citet{larson81} finds
that the internal velocity dispersion of GMCs, $\sigma$, scales with
their size, $R_{\mathrm{GMC}}$, as $\sigma \propto
R_{\mathrm{GMC}}^{0.38}$.  Assuming the GMCs are in virial
equilibrium, this leads to a mass-radius relation of the form
$R_{\mathrm{GMC}} \propto M_{\mathrm{GMC}}^{0.57}$.  Also assuming
virial equilibrium, \citet{solomon87} find
$R_{\mathrm{GMC}}~=~1/540\cdot M_{\mathrm{GMC}}^{0.5}$.  
These observations are consistent with GMCs having a constant surface
density ($\Sigma \equiv M_{\mathrm{GMC}}/\pi R_{\mathrm{GMC}}^{2}$).

From the Ebert-Bonnor relations for pressure bounded,
self-gravitating, isothermal spheres \citep{ebert55, bonnor56}, both
Eq.~\ref{eq:Ebert-Bonnor} as well as:
\begin{equation}
\Sigma \propto P_{s}^{1/2}
\end{equation}
can be derived, in which $P_{s}$ is the surface pressure
\citep{harris94, ashman01}.  So the observed mass-radius relation and
constant surface density for clouds are expected if the surface
pressure is constant \citep{elmegreen89}.

When clusters emerge from GMCs, the mass-radius relation appears to be
erased, indicating that high-mass clusters have higher stellar
densities than low-mass clusters.  A \emph{constant} stellar density
would predict $\reff \propto M_{\mathrm{cl}}^{1/3}$, which is not
observed.  \citet{zepf99} find $\reff \propto L^{0.07}$ for young
clusters in NGC~3256, where $L$ is the luminosity of the cluster which
scales directly with the mass (since their cluster sample suggests
that colour is independent of luminosity and therefore they assume
that the mass-to-light ratio is mostly independent on luminosity).
For clusters in a sample of (non-interacting) spiral galaxies,
\citet{larsen04b} finds $\reff \propto M_{\mathrm{cl}}^{0.10\pm
0.03}$.  The effective radius of the old Galactic GCs also does not
seem to correlate with their luminosity and thus their mass
\citep{vandenbergh91}.  \citet{mackey03a} report that for a sample of
53 rich LMC clusters, there seems to be no strong correlation between
their mass and core radius, either.

\begin{figure}
\centering
 \includegraphics[width=85mm]{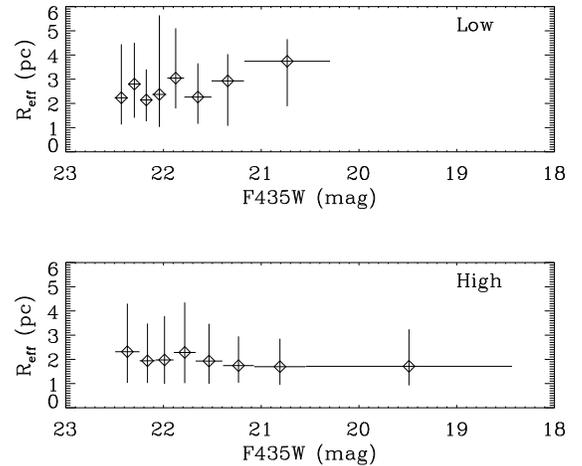}
\caption{The median \reff\ versus \bband\ magnitude for our sample of
317 clusters in the low background region (top) and for our sample of
728 clusters in the high background region.  The bins contain an equal
number of clusters and the vertical error bars indicate the range in
which 68\% of the clusters is contained.}
\label{fig:m51_magnitude_vs_radius}
\end{figure}

In Fig.~\ref{fig:m51_magnitude_vs_radius} we show \reff\ versus
magnitude in \bband\ for the M51 clusters, split in the low and
high background region.  For these clusters we do not have mass
estimates.  However, it is expected that the age range for the largest
fraction of this cluster sample is not very large, because most
clusters are blue and located in the spiral arms.  Many clusters are
therefore expected to have similar mass-to-light ratios and therefore
any mass-radius relation should also be visible as a relation between
magnitude and radius.  Fig.~\ref{fig:m51_magnitude_vs_radius} shows
that clusters in the high background regions show a slight trend of
radius decreasing with luminosity.  Clusters in the low background
regions show a less obvious trend, although the median \reff\ of the
brightest bin is considerably larger, especially compared to the
brightest bin in the high background region.  A fit on the unbinned
data points of the form $\reff \propto L^{x}$, with $L$ the luminosity
in the \bband\ passband, resulted in $x = 0.15\pm0.02$ and $x =
-0.11\pm0.01$ for the low and high background region, respectively.
We verified that applying the size-dependent aperture correction of
Eq.~\ref{eq:AC}, instead of the constant aperture correction for a
3~pc source (\S~\ref{subsec:Photometry}), would not change this result
considerably ($x = 0.13\pm0.02$ and $x = -0.10\pm0.01$ for the low and
high background, respectively).

It is \emph{not} likely that this observed differences in power-law
indices is a bias due to our detection limits, since we use magnitude
limits (\S~\ref{subsec:Selection criteria}) brighter than the 90\%
completeness limits for $\sim$8~pc sources in the high background
regions.  Due to the lack of age estimates of these clusters, there is
a degeneracy between age and mass.  Therefore it is not certain if any
evolution in \reff\ with luminosity is mainly caused by age effects,
mass effects, or a combination of both. More measurements of the ages and masses of
clusters which also have accurate radii estimates are necessary to
break this age/mass degeneracy (e.g.\ through additional $U$-band
imaging).

\begin{figure}
\centering
 \includegraphics[width=85mm]{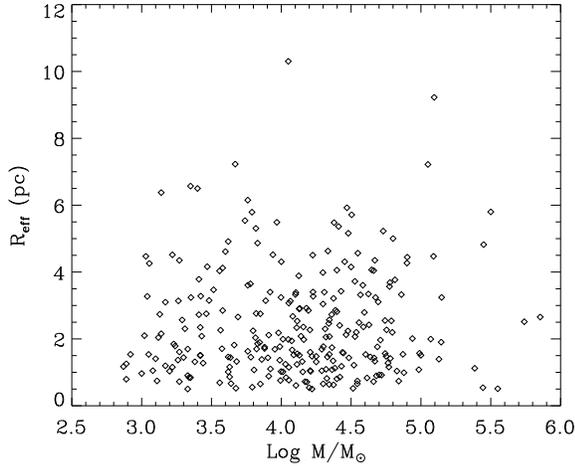}
\caption{\reff\ versus mass for the 271 clusters that were matched with the cluster sample of B05.}
\label{fig:m51_mass_vs_radius}
\end{figure}

Using the 271 clusters that were matched with the cluster sample of
B05 and for which we therefore have mass estimates, we show \reff\
versus mass in Fig.~\ref{fig:m51_mass_vs_radius}.  No apparent
relation is visible.  This sample is too small to make a distinction
between clusters in low and high background regions, since it mostly
covers the inner high background regions of M51.

We conclude that \emph{we do not find evidence for any direct relation
between mass and radius of the clusters}, although we find weak
relations between luminosity and radius, changing with background
region.  The suggested explanations for a lacking mass-radius
relation were already mentioned in \S~\ref{subsec:The slope of the
radius distribution}, but we stress again that none of these scenarios
are currently capable of explaining the observed differences in the
radius distributions between clouds and clusters.

\section{Summary and conclusions}
  \label{sec:Summary and conclusions}

We have used the \hst\ \acs\ mosaic image of M51 to detect 7698, 6846
and 5024 stellar clusters across the spiral disc in \bband, \vband\
and \iband\, respectively, based on effective radius (\reff)
measurements.  We presented the dataset and described the methods used
to select our cluster sample, including tests with artificial clusters
to show the accuracy, limits and robustness of our methods.  We
divided the data in 3 regions with respectively a low, intermediate
and high background, where the high background traces the spiral arms.
We selected a sample of 1284 clusters with the most accurate radius
estimates to study the radius distribution and relations between
radius, mass, luminosity, galactocentric distance and background
region.  From these studies we conclude the following:

\begin{enumerate}
\item The effective radii of the clusters are distributed between our
fixed lower limit of 0.5~pc and $\sim$10~pc (Fig.~\ref{fig:Radius
distribution linear}). The mean and median \reff\ of our accurate
radii sample are 2.5 and 2.1~pc, respectively.  This is smaller than
what is generally found for young clusters in spiral galaxies.

\item The radius distribution of clusters in M51 can not be fitted
with a power law similar as the one for star-forming gas clouds.  This
suggests that shortly after the formation of the clusters from a
fractal gas, their radii have have changed in a non-uniform way.

\item 70 clusters in our sample satisfy the colour criteria for being
old GCs.  These clusters are slightly larger than the rest of the
cluster sample (median $\reff = 2.4$~pc).  We find 6 clusters in our
sample satisfying the criteria for being ``faint fuzzy'' star clusters
projected onto the disc of M51 (Fig.~\ref{fig:FF}).

\item The largest cluster in our sample has $\reff = 21.6$~pc and a
projected galactocentric distance of 1.02~kpc (Fig.~\ref{fig:212995}).
Assuming low extinction and metallicity ($E(B-V)<
0.1$, $Z=$~0.2--1.0~$\mathrm{Z_{\sun}}$), we estimate its age to be
$2.7^{+2.8}_{-1.8}$~Gyr and its mass to be
$2.5^{+3.3}_{-1.1}$$\times10^{5}~\mathrm{M_{\sun}}$.  Assuming
extremely low metallicity ($0.02~\mathrm{Z_{\sun}}$) results in
$13.2^{+2.8}_{-9.7}$~Gyr and
$8.0^{+3.5}_{-4.6}$$\times10^{5}~\mathrm{M_{\sun}}$, for its age and
mass, respectively.

\item Comparing clusters in the low and high background regions, we
find that the high background regions, i.e.\ the spiral arms, have a
higher fraction of blue clusters, consistent with the idea that these
regions are the preferred formation sites for clusters
(Fig.~\ref{fig:Background_vs_color}).

\item We detect an increase in \reff\ with colour, most strongly for
\bv.  Since we detect most of the redder clusters outside the spiral
arms, the median \reff\ outside the spiral arms is larger than inside
the spiral arms: 2.7 and 1.9~pc, respectively. The radius distribution
of clusters in the low background region also shows a more pronounced
turnover around 1.5~pc (Fig.~\ref{fig:Radius
distribution linear}).  We speculate that if the observed spread in
colour is also a spread in the ages of the clusters, this observation
suggests a dynamical expansion of the clusters with age.  In this
hypothesis, newly formed clusters in spiral arms are generally small,
their radii reflecting the high surrounding pressure of the parental
gas clouds.  In the subsequent early evolution of the clusters an
increase in size is expected, likely due to dynamical heating from
cluster-cluster and cluster-cloud encounters and due to the removal of
binding energy when the clusters lose mass.

\item We do not observe a strong correlation between \reff\ and
galactocentric distance for the clusters in the disc of M51 out to
$\sim$13~kpc.  A weak trend is visible of the form $R_{\mathrm{eff}}
\propto R_{\mathrm{G}}^{0.12\pm0.02}$, but the scatter is large. For
old GCs, mainly residing in the outer regions or halo of other
galaxies, a steeper relation is observed, possibly caused by the
decreasing pressure of their parental gas clouds with galactocentric
distance.  The weaker relation for the clusters in M51 could be
explained by the observation that most of the clusters reside in the
spiral arms.  Since the spiral arms are expected to have a higher
pressure and they extend out to large galactocentric distances, a
strong radius-distance correlation is not expected.
  
\item We do not observe a correlation between cluster mass and radius
for the 271 clusters of which we have mass estimates.  We find weak
relations between cluster luminosity and radius for our sample of 1284
clusters.  If fitted with a power law of the form $\reff \propto
L^{x}$, we find $x=0.15\pm0.02$ and $x = -0.11\pm0.01$ for the low and
high background region, respectively
(Fig.~\ref{fig:m51_magnitude_vs_radius}). Explanations of the lack of
a strong mass-radius or luminosity-radius relation probably need to be
sought in the early dynamical evolution (expansion) of the clusters
just after their formation.  Current scenarios which focus on the
expansion of clusters due to either dynamical heating or the removal
of binding energy due to gas expulsion are not consistent with the
observed differences in the radius distributions between clouds and
clusters.

\end{enumerate}

\begin{acknowledgements}
We thank Peter Anders for useful discussions, tips and
comments.  We also thank Marcelo Mora at ESO/Garching for discussions
and for kindly providing us with the empirical PSFs. We would like to
thank Max Mutchler, Richard Hook and Andrew Fruchter for discussions
regarding the effects of the drizzle routine on the PSF. We thank Narae Hwang
for kindly providing us the list of faint fuzzies around NGC~5195. 
\end{acknowledgements}

\bibliographystyle{aa} 
\bibliography{7511}


\end{document}